\documentclass[11pt]{article}
\usepackage{amsmath,amssymb}%
\usepackage{pdflscape}
\usepackage{color}

\usepackage{graphicx}  

\usepackage{color}

\usepackage{psfrag}
\usepackage{epsfig}
\usepackage{epstopdf}
\usepackage{amssymb,latexsym}
\usepackage[mathscr]{eucal}
\usepackage{cite}
\usepackage{url}
\usepackage{ntheorem}

\DeclareFontFamily{OT1}{rsfs}{} \DeclareFontShape{OT1}{rsfs}{m}{n}{
<-7> rsfs5 <7-10> rsfs7 <10-> rsfs10}{}
\DeclareMathAlphabet{\mycal}{OT1}{rsfs}{m}{n}

\newcommand{\tilr}{\tilde r}%

\newcommand{\jlcax}[1]{}
\newcommand{\eean}{\nonumber\end{eqnarray}}

\newcommand{\U}{{\bf U}}

\newcommand{\ttheta}{\tilde\theta}

\newcommand{\kk}[1]{}

\newcommand{\mcH}{{\mycal H}}

\newcommand{\beq}{\begin{equation}}

\newcommand{\tx}{x}%

\newcommand{\FS}       
                  {F}

\newcommand{\HS} 
       {H_{\mbox{\scriptsize volume}}}

{\ptc{this should be removed in the oberwolfach version}}%

\newcommand{\mcA}{{\mycal A}}%
\newcommand{\eeal}[1]{\label{#1}\end{eqnarray}}
\newcommand{\bed}{\begin{deqarr}}
\newcommand{\eed}{\end{deqarr}}
\newcommand{\bedl}[1]{\begin{deqarr}\label{#1}}
\newcommand{\eedl}[2]{\arrlabel{#1}\label{#2}\end{deqarr}}

\newcommand{\bel}[1]{\begin{equation}\label{#1}}
\newcommand{\bea}{\begin{eqnarray}}
\newcommand{\bean}{\begin{eqnarray}\nonumber}
\newcommand{\beal}[1]{\begin{eqnarray}\label{#1}}
\newcommand{\eea}{\end{eqnarray}}

\newcommand{\be}{\begin{equation}}
\newcommand{\eeq}{\end{equation}}
\newcommand{\ee}{\end{equation}}
\newcommand{\beqa}{\begin{eqnarray}}
\newcommand{\eeqa}{\end{eqnarray}}
\newcommand{\beqan}{\begin{eqnarray*}}
\newcommand{\eeqan}{\end{eqnarray*}}
\newcommand{\ba}{\begin{array}}
\newcommand{\ea}{\end{array}}

\newcommand{\mcM}{{\mycal M}}

\newcommand{\mnote}[1]
{\protect{\stepcounter{mnotecount}}$^{\mbox{\footnotesize
$
\bullet$\themnotecount}}$ \marginpar{
\raggedright\tiny\em
$\!\!\!\!\!\!\,\bullet$\themnotecount: #1} }

\newcommand{\warn}[1]
{\protect{\stepcounter{mnotecount}}$^{\mbox{\footnotesize
$
\bullet$\themnotecount}}$ \marginpar{
\raggedright\tiny\em
$\!\!\!\!\!\!\,\bullet$\themnotecount: {\bf Warning:} #1} }

\newcommand{\R}{\mathbb R}

\newcommand{\eq}[1]{(\ref{#1})}

\newcommand{\ptc}[1]{\mnote{{\bf ptc:}#1}}

\newcommand{\diag}{\mbox{\rm diag}}

\newcommand{\beqar}{\begin{deqarr}}
\newcommand{\eeqar}{\end{deqarr}}

\newcommand{\beaa}{\begin{eqnarray*}}
\newcommand{\eeaa}{\end{eqnarray*}}

\newcommand{\trho}{{\tilde \rho}}

\newcommand{\ty}{{\tilde y}}
\newcommand{\hrho}{\hat\rho}

\newcommand{\hx}{{\hat x}}
\newcommand{\hy}{{\hat y}}

\newcommand{\levoca}[1]{\ptc{#1}}

\renewcommand{\tx}{\tilde x}
\renewcommand{\levoca}[1]{}

\newcommand{\cpm}{c_\pm^2}%
\newcommand{\csm}{c_-^2}%
\newcommand{\csp}{c_+^2}%

{\catcode `\@=11 \global\let\AddToReset=\@addtoreset}
\AddToReset{equation}{section}

{\catcode `\@=11 \global\let\AddToReset=\@addtoreset}
\AddToReset{figure}{section}

{\catcode `\@=11 \global\let\AddToReset=\@addtoreset}
\AddToReset{table}{section}

\newcounter{mnotecount}[section]

\renewcommand{\themnotecount}{\thesection.\arabic{mnotecount}}

\newcommand{\ednote}[1]{}

\definecolor{bluem}{rgb}{0,0,0.5}

\definecolor{mycolor}{cmyk}{0.5,0.1,0.5,0}
\definecolor{michel}{rgb}{0.5,0.9,0.9}

\definecolor{turquoise}{rgb}{0.25,0.8,0.7}
\definecolor{bluem}{rgb}{0,0,0.5}

\definecolor{MDB}{rgb}{0,0.08,0.45}
\definecolor{MyDarkBlue}{rgb}{0,0.08,0.45}

\definecolor{MLM}{cmyk}{0.1,0.8,0,0.1}
\definecolor{MyLightMagenta}{cmyk}{0.1,0.8,0,0.1}

\definecolor{HP}{rgb}{1,0.09,0.58}
\definecolor{dkgreen}{rgb}{0,0.8,0}

\newcommand{\ri}{R_i}
\newcommand{\rj}{R_j}

\newcommand{\ptclater}[1]{\mnote{ {\bf ptc (to do later):}{\sc {#1}} \ }}

\begin{document}
\title{On smoothness of Black Saturns}

\author{Piotr T.~Chru\'sciel\thanks{PTC was supported in part  by the EC project KRAGEOMP-MTKD-CT-2006-042360,
by the Polish Ministry of Science and Higher Education grant Nr
N N201 372736, and by the EPSRC Science and Innovation award to
the Oxford Centre for Nonlinear PDE (EP/E035027/1).}
\\
Institute of Physics, University of Vienna \\
\\
Micha\l\ Eckstein\\
Instytut Fizyki, Uniwersytet Jagiello\'nski, Krak\'ow \\
\\
Sebastian J. Szybka\thanks{SSz was supported in part by the
Polish Ministry of Science and Higher Education grant Nr N N202
079235,
and by the Foundation for Polish Science.} \\
Obserwatorium Astronomiczne, Uniwersytet Jagiello\'nski, Krak\'ow}
\maketitle{}
\date{}
\begin{abstract}
We prove smoothness of the domain of outer communications
(d.o.c.) of the Black Saturn solutions of Elvang and Figueras.
We show that the metric on the d.o.c.\ extends smoothly across
two disjoint event horizons with topology $\R\times S^3$ and
$\R\times S^1\times S^2$. We establish stable causality of the d.o.c. when 
the Komar angular momentum of the spherical component of the horizon vanishes,
and present numerical evidence for stable causality in general.
\end{abstract}

\tableofcontents

\section{Introduction}
 \label{SI}

 \renewcommand{\ptc}[1]{}
\ptc{the command ptclater momentarily disabled}
\renewcommand{\ptclater}[1]{}
In~\cite{EF},
Elvang and Figueras introduced a family of vacuum
five-dimensional asymptotically flat metrics, to be found in
Appendix~\ref{ssMetrCoeff13VI}, and presented evidence that
these metrics describe two-component black holes, with Killing
horizon topology $\R\times\big( (S^1\times S^2) \cup
S^3)\big)$. In this paper we construct  extensions of the
metrics across Killing horizons, with the Killing horizon
becoming an event horizon in the extended space-time. Now, it
is by no means clear that those metrics have no singularities
within their domains of outer communications (d.o.c.), and the
main purpose of this work is to establish this for non-extreme
configurations. Again, it is by no means clear that the
d.o.c.'s of the solutions are well behaved causally. We prove
that those d.o.c.'s  are stably causal when the parameter $c_2$
vanishes  (this condition is equivalent to the vanishing of the
Komar angular momentum of the spherical component of the
horizon, compare \cite[Equation~(3.39)]{EF}), and present
numerical evidence suggesting that this is true in general.

Given the analytical and numerical evidence presented here, it
appears that the Black Saturn metrics describe indeed well
behaved black hole space-times within the whole range of
parameters given by Elvang and Figueras, except possibly for
the degenerate cases when some parameters $a_i$ coalesce, a
study of which is left for future work.
 \ptclater{}%
In particular we have rigorously established that the Black
Saturn metrics with $c_2=0$ and with distinct $a_i$'s have a
reasonably well behaved neighbourhood of the d.o.c. Our
reticence here is related to the fact that we have not proved
global hyperbolicity of the d.o.c., which is often viewed as a
desirable property of the domains of outer communications of
well behaved black holes. In view of our experience with the
Emperan-Reall metrics~\cite{CC}, the proof of global
hyperbolicity (likely to be true) appears to be a difficult
task.

We use the notation of \cite{EF}, and throughout this paper we
assume that the parameters $a_i$  occurring in the metric are
pairwise distinct, $a_i\ne a_j $ for $i \ne j$.

\section{Regularity at {\protect$\lowercase{z=a_1}$}, {\protect$\lowercase{\rho=0}$}, and the choice of {\protect$\lowercase{c_1}$}}
 \label{Schoicec_1}

We consider the metric coefficient $g_{tt}$ on the set
$\{\rho=0, z<a_1\}$. A {\sc Mathematica} calculation shows that
$g_{tt}$ is a rational function with denominator given by
\bea &  -\left(2 (a_3-a_1) (a_2-a_4)+(a_5-a_1) c_1 c_2\right)^2
(z-a_1) (z-a_2) (z-a_4  )
      \;,
\eeal{21VII.1x}
which clearly vanishes as $z$ approaches $a_1$ from below (we
will see in Section \ref{s6XII.1} that the first multiplicative
factor is non-zero with our choices of constants). On the other
hand, its numerator has the following limit as $z\to a_1$,
\bea (a_2-a_1)^2 (a_3-a_1) (a_5-a_1) \left(2 (a_3-a_1)
(a_4-a_1)-(a_5-a_1) c_1^2\right) c_2^2
 \;,
\eeal{21VII.1}
{which is non-zero unless $c_2$ vanishes or $c_1$ is chosen to
make  the before-last factor vanish}:
\bel{21VII.4}
 c_1=\pm\sqrt{\frac{2( a_3 - a_1  ) ( a_4 - a_1  )}{a_5  - a_1 }} \ne 0
 \;.
\ee
This coincides with Equation~(3.7) of~\cite{EF}.

By inspection, one finds that the metric is invariant under the
transformation
$$
 (c_1,c_2,\psi)\mapsto (-c_1,-c_2,-\psi)
 \;.
$$
Thus, an overall change of sign $(c_1,c_2)\mapsto (-c_1,-c_2)$
can be  implemented by a change of orientation of the angle
$\psi$. Hence, to understand the global structure of the
associated space-time, it suffices to consider the case
$$
 c_1 >0
\;;
$$
this will be assumed throughout the paper from now on.

If \eq{21VII.4} does not hold, the Lorentzian norm squared
$g_{tt}=g(\partial_t,\partial_t)$ of the Killing vector
$\partial_t$ is unbounded as one approaches $a_1$; a well known
argument shows that this leads to a geometric singularity.

We show in Section~\ref{ssAor25III10} that the \emph{choice
\eqref{21VII.4} is necessary for regularity of the metric}
regardless of whether or not $c_2=0$: without this choice,
$g_{\psi\psi}$ would be unbounded near $a_1$, leading to a
geometric singularity as before.

With the choice \eqref{21VII.4}  of $c_1$, or with $c_2=0$, the
point $\alpha_1:=(\rho=0,z=a_1)$ in the quotient of the
space-time by the action of the isometry group becomes a
\emph{ghost point}, in the sense that it has no natural
geometric interpretation, such as a fixed point of the action,
or the end-point of an event horizon. Now, the functions
$$
 R_i := \sqrt{\rho^2 + (z-a_i)^2}
$$
are \emph{not} differentiable at $\rho=0,z=a_i$. So, a generic
function of $R_1$ will have some derivatives blowing up at
$\rho=0,z=a_1$. However, this will not happen for functions
which are smooth functions of $R_1^2$. It came as a major
surprise to us that the choice of $c_1$ above, determined by
requiring boundedness of $g_{tt}$ on the axis near $a_1$, also
leads to \emph{smoothness} of all metric functions near
$z=a_1$. It turns out that there is a general mechanism which
guarantees that; this will be discussed
elsewhere~\cite{ChNguyenRemSing}.

To establish that the metric is indeed smooth near the ghost
point $\alpha_1$, we start with
$$
 g_{tt} = -\frac {H_y}{H_x}=-\frac { FH_y}{F H_x}=:\Phi(\mu_1,\mu_A,c_1,c_2,\rho^2)
 \;,
 $$
where $A$ runs from two to five. $\Phi$ is a rational function
of its arguments, and hence a rational function of $R_1$. So
$g_{tt}$ will be a smooth function of $R_1^2$ near $R_1=0$ if
and only if $\Phi$ is even in $R_1$:
\bel{2XII.1}
 \Phi(R_1-(z-a_1),\mu_A,c_1,c_2,\rho^2)=
 \Phi(-R_1-(z-a_1),\mu_A,c_1,c_2,\rho^2)
 \;,
\ee
assuming moreover that the right value of $c_1$ has been
inserted. (We emphasise that neither $FH_x$ or $FH_y$ are even
in $R_1^2$, so
there is a non-trivial factorisation involved;%
\footnote{We are grateful to H.~Elvang and P.~Figueras for
drawing our attention to the fact that this factorisation takes
place in the Emparan-Reall limit of the Black Saturn metric.}
moreover $g_{tt}$ is \emph{not} even in $R_1$ for arbitrary
values of the $c_i$'s, as is seen by setting $c_1=c_2=0$.) Now,
there is little hope of checking this identity by hand after
all functions have been expressed in terms of $\rho$, $z$, and
the $a_i$'s, and we have not been able to coerce Mathematica to
deliver the required result in this way either. Instead, to
avoid introducing new functions or parameters into $\Phi$, we
first note that
$$
 -R_1 - (z-a_1) = - \frac{\rho^2}{\mu_1}
 \;,
$$
and so \eq{2XII.1} reads
$$
\Phi(\mu_1,\mu_A,c_1,c_2,\rho^2)= \Phi\big(- \frac{\rho^2}{\mu_1},\mu_A,c_1,c_2,\rho^2\big)
 \;.
 $$
From the explicit form of the functions $FH_x$ and $FH_y$ we
can write
$$
\Phi(\mu_1,\mu_A,c_1,c_2)- \Phi\big(-
\frac{\rho^2}{\mu_1},\mu_A,c_1,c_2\big) = \frac{\sum_{i=0}^{4} \Phi_i (c_1c_2)^i}{G}
 \;,
$$
where the $\Phi_i$'s are polynomials in $c_1^2$, $\mu_i$ and
$\rho^2$, and $G$ is a polynomial in $\mu_i$, $c_1$, $c_2$ and
$\rho^2$. One then checks with {\sc Mathematica} that each of
the coefficients $\Phi_i$  has a multiplicative factor that
vanishes after applying the identity \eq{21IV.42} below to
replace each occurrence of $c_1^2$ in terms of the $\mu_i$'s:
$$
 c_1^2=\frac{(-\mu_1+\mu_3) (-\mu_1+\mu_4) \mu_5 (\mu_1 \mu_3+\rho^2) (\mu_1 \mu_4+\rho^2)}
 {\mu_1 \mu_3 \mu_4 (-\mu_1+\mu_5) (\mu_1 \mu_5+\rho^2)}
 \;.
$$
It is rather fortunate that each of those coefficients has a
vanishing factor, as we have not been able to convince {\sc
Mathematica} to carry out a brute-force calculation on all
coefficients at once.

An identical analysis applies to $g_{\rho\rho}=g_{zz}$ and
$\omega_\psi/H_y$; regularity of $g_{\psi\psi}$ immediately
follows; there is nothing to do for $g_{\varphi\varphi}$.
Before doing these calculation, care has to be taken to
eliminate, with the right signs, all square roots of squares
that appear in the definition of $\omega_\psi$.

\section{Asymptotics at infinity: the choice of \protect$\lowercase{q}$ and \protect$\lowercase{k}$}
 \label{ssAatinf13VII}

We wish to check that the Black Saturn metric is asymptotically
flat. As a guiding principle,  the Minkowski metric on $\R^5$
is written in coordinates adapted to $\U(1)\times \U(1)$
symmetry as
%
\bean
 \eta & = & -dt^2 + d\tx^2 + d\ty^2 +d\hx^2 + d\hy^2
\\
 & = & -dt^2 + d\trho^2 + \trho^2 d\psi^2 + d\hrho^2 + \hrho^2 d\varphi^2
 \;,
\eeal{13VII.2}
with
$$
 (\tx,\ty)= \trho(\cos \psi, \sin \psi)\;,
 \quad
 (\hx,\hy)=
 \hrho(\cos \varphi, \sin \varphi)
  \;.
$$
Introducing $\rho$ and $\theta$ as polar coordinates in the
$(\hrho, \trho)$ plane,
$$
 (\hrho,\trho) = r (\cos \theta, \sin \theta)
 \;,
$$
the metric \eq{13VII.2} becomes
\bel{13VII.3}
 \eta = -dt^2 + dr^2 + r^2 d\theta^2 + r^2 \sin^2 \theta \, d\psi^2 + r^2 \cos^2 \theta \, d\varphi^2
 \;.
\ee
Note that $\theta\in[0,\pi/2]$ since both $\trho$ and $\hrho$
are positive in our range of interest.

As outlined by Elvang and Figueras in~\cite{EF}, relating the
$(\rho,z,\psi,\varphi)$ coordinates of the Black Saturn metric
to the $(r,\theta,\psi,\varphi)$ coordinates of \eq{13VII.3}
via the formulae
\bel{13VII.4}
 \rho = \frac 12 r^2 \sin 2 \theta\;,\quad z= \frac 12 r^2 \cos 2 \theta
  \;, \quad \theta \in \left[0,\frac \pi 2\right]
 \;,
\ee
should lead to a metric which is asymptotically flat. Under
\eq{13VII.4} the metric \eq{13VII.3} becomes
\bel{13VII.5}
 \eta = -dt^2 +  r^{-2} (d\rho^2 + dz^2)  + r^2 \sin^2 \theta \, d\psi^2 + r^2 \cos^2 \theta \, d\varphi^2
 \;,
\ee
so that in such coordinates a set of necessary conditions for
asymptotic flatness reads
\beal{13VII.6}
 & g_{tt}\to -1\;,\
 r^{-1}\sin^{-1} \theta \, g_{t\psi}\to 0\;,\
 &
\\
 &
 r^2 g_{\rho \rho}=
 r^2 g_{zz }\to 1\;, \
 r^{-2}\sin^{-2}\theta \,g_{\psi \psi}\to 1\;, \
 r^{-2}\cos^{-2}\theta \,g_{\varphi \varphi}\to 1
 \;,
 &
\eeal{13VII.6b}
when $r$ tends to infinity. One also needs to check that all
metric components are suitably behaved when transformed to the
coordinates $(\tx,\ty,\hx,\hy)$ above. Finally, each derivative
of any metric components should decay one order faster than the
preceding one.

We start by noting that
$$
 z= \frac 12 r^2 (\cos^2 \theta - \sin^2 \theta) = \frac 12 (\hrho^2 - \trho^2)
$$
which is a smooth function of $(\tx, \ty, \hx, \hy)$. On the
other hand,
$$ \rho = r^2 \sin \theta\, \cos \theta = \hrho \trho
$$
is \emph{not} smooth, but its square is. This implies that all
the functions appearing in the metric are smooth functions of
$(\tx, \ty, \hx, \hy)$, except perhaps at zeros of the
functions $R_i$ and of the denominators; the former clearly do
not occur at sufficiently large distances, while the
denominators have no zeros for $\rho >0$ by
Section~\ref{SRegularity13VI}, and at $\rho=0$ away from the
points $a_i$ by Sections~\ref{ssAor13VII} and
\ref{ssAor25III10}.

To control the asymptotics we note that $\mu_i=O(r^2)$, but
more precise control is needed. Setting $R^2:=\rho^2+z^2
=r^4/4$, a Taylor expansion within the square root gives
\beaa
 \mu_i & = & \sqrt{\rho^2 + (z-a_i)^2} - (z-a_i)
\\& = & R\sqrt{1 -  \frac{2z a_i - a_i^2}{R^2}} - (z- {a_i})
\\
  & = &
   \big(r^2 + 2a_i + 2\frac {a_i^2}{r^2}(1+\cos 2\theta)
\big)\sin^ 2\theta + O(r^{-4})
\\
 & = & (r^2 + 2a_i) \sin^ 2\theta + O(r^{-2})\;.
\eeaa
For $z\le 0$ this can be rewritten as
\bel{13VII.1}
 \mu_i
 =
  (r^2 + 2a_i + O(r^{-2})) \sin^ 2\theta\;.
\ee
To see that the last equation remains valid for $z \ge 0$ we
write instead
\beaa
 \mu_i & = & \frac{\rho^2}{\sqrt{\rho^2 + (z-a_i)^2} + (z-a_i)}
\\
 & = & \frac{R^2 \sin^ 22\theta}{R\sqrt{1 -  \frac{2z a_i - a_i^2}{R^2}} + (z-
 {a_i})}
\\
 & = & \frac{R  \sin^ 22\theta}{ 1 -  \frac{z a_i}{R^2}+ \frac z R-
\frac {a_i}R + O(R^{-2})}
\\
 & = & \frac{R  \sin^ 22\theta}{(1+ \frac z R)( 1 -  \frac{ a_i}{R}+ O(R^{-2}))}
\\
  & = &
   \big(R + a_i
 + O(R^{-1}) \big)\frac{\sin^ 2 2\theta}{1+ \cos 2 \theta}
 \;,
\eeaa
and we have recovered \eq{13VII.1} for all $z$, for $r$ large,
uniformly in $\theta$.

The above shows that $\mu_i-\mu_j =O(1)$ for large $r$; in
fact, for $i\ne j$,
$$
 \mu_i - \mu_j =\big(2 (a_i - a_j) + O(r^{-2})\big)\sin^2 \theta \;.
$$
Keeping in mind that
$$
\rho^2+ \mu_i \mu_j \approx r^4 \sin^2 \theta
 \;,
$$
where we use $f\approx g$ to denote that $C^{-1} \le f/g \le C$
for large $r$, for some positive constant $C$, we are led to
the following uniform estimates
%
\beaa
 &
 M_0 \approx r^{30}\sin^{26} \theta
  \;,
 &
\\
&
 M_1  \approx r^{24}\sin^{28} \theta\sin^{2} 2\theta\;,\
 M_1 \frac{\rho^2}{\mu_1\mu_2}  \approx r^{24}\sin^{24} \theta\sin^{4} 2\theta\;,
&
\\
 &
 M_2  \approx r^{28}\sin^{24} \theta\sin^{2} 2\theta\;,\
 M_2   \frac{\mu_1\mu_2}{\rho^2} \approx r^{28}\sin^{28} \theta
  \;,\
 &
\\
&  M_3  \approx r^{30}\sin^{26}\theta\
 \;,\
 M_4  \approx r^{30}\sin^{26}  \theta
 \;,
 &
\\
&
  F \approx r^{48}\sin^{34}  \theta
  \;,
 &
\\
 &
  \displaystyle
   G_x = \frac{r^2\sin^2 2\theta}{4 \sin^2 \theta}\big(1+O(r^{-2})\big)\approx r^2 \cos^2 \theta\; ,
  &
\\
  &
  P =  (\mu_3\, \mu_4+ \rho^2)^2
      (\mu_1\, \mu_5+ \rho^2)
      (\mu_4\, \mu_5+ \rho^2)\approx r^{16}\sin^8\theta\, .
 &
\eeaa
This shows that, for large $r$,
\beaa
   H_x &=& F^{-1} \,
   \bigg[ \underbrace{M_0 +
    c_1\, c_2\, M_3 + c_1^2 c_2^2\, M_4}_{\approx r^{30} \sin^{26}\theta} +O(r^{28}\sin^{28}\theta)
   \bigg] \, , 
   \\[2mm]
\nonumber    H_y &=& F^{-1} \,
   \frac{\mu_3}{\mu_4}\,
   \bigg[\underbrace{ M_0 \frac{\mu_1}{\mu_2}
   +  c_1\, c_2\, M_3
   + c_1^2 c_2^2\, M_4 \frac{\mu_2}{\mu_1}}_{\approx r^{30} \sin^{26}\theta} +O(r^{28}\sin^{28}\theta)\bigg] \;,
\eeaa
and in fact the ratio tends to $1$ at infinity. We conclude
that
$$
 g_{tt}+1 = O(r^{-2})\;,
$$
uniformly in angles.

In order to check the derivative estimates required for the
usual notion of asymptotic flatness, we note the formulae
\begin{eqnarray*}
 \mu_i &=& a_i + 1/2 \bigg(-\hx^2 - \hy^2 + \tx^2 + \ty^2
\\
 && + \sqrt{
    4 a_i^2 -
     4 a_i (\hx^2 + \hy^2 - \tx^2 - \ty^2) + (\hx^2 + \hy^2 + \tx^2 +
     \ty^2)^2}
    \bigg)
    \;,
\\
 \rho^2 &= & (\hx^2 + \hy^2) (\tx^2 + \ty^2)
 \;.
\end{eqnarray*}
Since the $\mu_i$'s and $\rho^2$ are smooth functions at
sufficiently large distances,  it should be clear that every
derivative of any metric function decays one power of
$\sqrt{\hx^2 + \hy^2 + \tx^2 +
     \ty^2}$ faster than the immediately preceding one, as required.

The constant $q$ appearing in the metric is determined by
requiring that $g_{t\psi}\to 0$ as $r$ tends to infinity.
Equivalently, since $g_{tt}\to -1$,
$$
 q = - \lim_{r\to \infty} \frac {\omega_\psi}{H_y}
 \;.
$$
Now,
\beaa
  -\frac{\omega_\psi}{H_y}
 & = &
  - 2 \frac{
     c_1\, R_1\, \sqrt{M_0 M_1}
    -c_2\, R_2\, \sqrt{M_0 M_2}
    +c_1^2\,c_2\, R_2\, \sqrt{M_1 M_4}
    -c_1\,c_2^2\, R_1\, \sqrt{M_2 M_4}
  }
  {F H_y \sqrt{G_x}}
\\
  &=&
  2 c_2\frac {\mu_4}{\mu_3}\frac{
    \, R_2\, \sqrt{M_0 M_2}
    +c_1\,c_2\, R_1\, \sqrt{M_2 M_4}
    +O(r^{29})
  }
  { \sqrt{G_x}\big( M_0 \frac{\mu_1}{\mu_2}
   +  c_1\, c_2\, M_3
   + c_1^2 c_2^2\, M_4 \frac{\mu_2}{\mu_1} +O(r^{28})\big)}
  \; ,
\eeaa
where we have not indicated the angular dependence of the
subleading terms, but it is easy to check that the terms kept
dominate likewise near the axes. A {\sc Mathematica}
calculation gives
$$
 q=\frac{2 c_2\kappa_1}{
 2\kappa_1 - 2\kappa_1\kappa_2 + c_1 c_2\kappa_3}
 \;,
$$
which can be seen to be consistent with~\cite{EF}, when the
required values of the $c_a$'s are inserted.

In view of \eq{13VII.6b}, the constant $k>0$ needs to be chosen
so that
$$
 k^2 \lim_{r\to \infty} r^2 H_x P = 1
 \;.
$$
One finds
%
$$k^2=\frac{4\kappa_1^2 (-1 +\kappa_2)^2}{{(-2\kappa_1 (-1 +\kappa_2) +
    c_1 c_2\kappa_3)^2}}
    \;,
$$
as in~\cite{EF}. From \eq{13VII.1} and from what has been said
so far one immediately finds
\beaa
 \lim_{r\to \infty} r^{-2}\sin^{-2}\theta \,g_{\psi \psi}
  & = &  \lim_{r\to \infty}\frac {H_x G_y}{ r^2 \sin^2\theta H_y }
\\
  & = &    \lim_{r\to \infty}\frac {  G_y}{  r^2 \sin^2\theta}
  =   \lim_{r\to \infty}\frac {\mu_3 \mu_5}{  r^2 \sin^2\theta\, \mu_4}
\\
  & = &     1
 \;,
\eeaa
as desired. Finally, it is straightforward that
\beaa
 \lim_{r\to \infty} r^{-2}\cos^{-2}\theta \,g_{\varphi \varphi}
  &= &
    \lim_{r\to \infty} \frac {G_x  }{  r^2 \cos^2\theta} =
    \lim_{r\to \infty} \frac {\rho^2 \mu_4  }{  r^2 \cos^2\theta\, \mu_3 \mu_5}
\\
  & = &     1
 \;.
\eeaa
Further derivative estimates follow as before, and thus we have
proved:
\bel{13VII.11}
 g_{\mu\nu} - \eta_{\mu\nu} = O(r^{-2})\;, \quad
  \partial_{i_1}\ldots \partial_{i_ \ell}  g_{\mu\nu}   = O(r^{-2-\ell})
  \;.
\ee

 \section{Conical singularities and the choice of \protect$\lowercase{c_2}$}
 \label{s6XII.1}

It is seen in Table~\ref{T18VII.1} below that $g_{\varphi
\varphi}$ vanishes for $\{z\le a_5\}\cup\{a_4 <z\le a_3\}$,
while $g_{\rho \rho}$ does not, which implies that the set
$\{z< a_5\}\cup\{a_4 <z < a_3\}$ is an axis of rotation for
$\partial_\varphi$. In such cases the ratio
$$
 \lim_{\rho \to 0} \frac{\rho^2 g_{\rho\rho}}{g_{\varphi \varphi}}
$$
determines the periodicity
of $\varphi$ needed to avoid a conical singularity at zeros of
$\partial_\varphi$, and thus this ratio should be constant
throughout this set. This leads to two equations. For $\{z\le
a_1\}$, the choice of $k$ already imposed by asymptotic
flatness leads to
\bel{15VIII.1}
 \lim_{\rho\rightarrow
0}\frac{g_{\rho\rho}}{g_{\varphi\varphi}} \rho^2=1
 \;.
\ee
Either by a direct calculation, or invoking analyticity at
$\rho=0$ across $z=a_5$, one finds  that the same limit is
obtained for $a_1 <z\le a_5$ with the choices of $k$ ad $c_1$
determined so far. The requirement that \eq{15VIII.1} holds as
well for $a_4 <z\le a_3$, together with the choice of $k$
already made, gives an equation that determines $c_2$:
\beaa \lefteqn{  \lim_{\rho\rightarrow
 0}\frac{g_{\rho\rho}}{g_{\varphi\varphi}}\rho^2=2 (a_2-a_1) (a_3-a_4) \times}
 &&
\\
&&   \sqrt{\frac{(a_3-a_1) (a_2-a_4)}{(a_2-a_5) (a_3-a_5) (2
(a_3-a_1) (a_2-a_4)+(a_5-a_1)
 c_1 c_2)^2}}
\\
 &&
 =1\;.
\eeaa
Therefore, to avoid a conical singularity one has to choose
\bea\label{c_2conical} c_2=2 \frac{(a_3-a_1) c_1 S_1\pm
(a_1-a_2) (a_3-a_4) S_2}{(a_1-a_5) (a_5-a_2) (a_5-a_3)
c_1^2}, \eea
where
\bea
\nonumber
S_1&=&(a_2-a_4) (a_2-a_5) (a_3-a_5)\;,\\
\nonumber
S_2&=&\sqrt{( a_3 -a_1) c_1^2 S_1}\;.
\eea
Equivalently,
%
%
$$
c_2=
\sqrt 2 (  a_4 - a_2  ) \frac{\pm(a_1  - a_2  ) (a_3   - a_4  ) +
   \sqrt{(a_1  - a_3  ) ( a_4 -a_2  ) (a_2   - a_5 ) (a_3   - a_5 )}}{\sqrt{(a_1  -
   a_4  ) (a_2   - a_4  ) (a_1  - a_5 ) (a_2   - a_5 ) (a_3   - a_5 )}}
 \;,
$$
as found in~\cite{EF}.

The case $c_2=0$, which arose in Section~\ref{Schoicec_1}, is
compatible with this equation for some ranges of parameters
$a_i$, we return to this question in
Section~\ref{ssAor25III10}.

It follows from the analysis of Section~\ref{ssAatinf13VII}
that the analogous regular-axis condition for $z>a_2$,
\bel{15VIII.1x}
 \lim_{\rho\rightarrow
0}\frac{g_{\rho\rho}}{g_{\psi\psi}} \rho^2=1
 \;,
\ee
is satisfied at sufficiently large distances when $k$ assumes
the value determined there. One checks by a direct calculation
(compare \eq{24III10.1})  that the left-hand  side of
\eqref{15VIII.1x} is constant on $(a_2,\infty)$, and smoothness
of the metric across $\{\rho=0,z\in(a_2,\infty)\}$ ensues.

\section{The analysis}
 \label{sTa_{1}3VII}

\subsection{The sign of the $\mu_i$'s}
 \label{ssSign13VI}

Straightforward algebra leads to the identity, for $i\ne j$,
\bel{21IV.42}
 a_i-a_k = \frac{(\mu_i-\mu_k)(\rho^2+\mu_i\mu_k)}{2\mu_i\mu_k}
 \;.
\ee
Since all the $\mu_i$'s are non-negative, vanishing only on a
subset of the axis
$$ \mcA:=\{\rho=0\}
 \;,
$$
we conclude that
\bel{21IV.45} \text{the $\mu_i-\mu_k$'s have the same sign as the $a_i-a_k$'s.}
\ee
Furthermore from \eq{21IV.42} we find
\bel{21IV.43}
 \kappa_i:= \frac{a_{i+2}-a_1}{a_2-a_1} = \frac{(\mu_{i+2}-\mu_1)(\rho^2+\mu_1\mu_{i+1})}{2\mu_1\mu_{i+2}
 (a_2-a_1)} >0
 \;.
\ee

We infer that the functions $M_\nu$, $\nu=0, \ldots, 4$ are
non-negative: indeed, this follows from the fact that the
$\mu_\nu$'s are non-negative, together with \eq{21IV.45}.

\subsection{Positivity of $H_x$ for $\rho>0$}
 \label{ssPHx}

We wish to show that $H_x$ is non-negative, vanishing at most
on the axis $\{\rho = 0\}$; note that by the analysis in
Section~\ref{ssAatinf13VII}, $H_x$ certainly vanishes at
$\theta=0$.

Now, $H_x$ vanishes if and only if its numerator vanishes:
\bel{Hx.zeros} M_0+c_1^2 M_1+c_1 M_3 c_2 + \big(M_2+c_1^2
M_4\big)c^2_2=0\;. \ee
This equation may be seen as a quadratic equation for $c_2$;
its discriminant
\beaa
 \Delta&=&c_1^2 M_3^2 - 4 (M_0 + c_1^2 M_1) (M_2 + c_1^2
M_4) \eeaa
can be brought, using {\sc Mathematica}, to the form
\bean
\Delta
&=&-4 (\mu_1-\mu_2)^2 \mu_2^2 \mu_3 (\mu_2-\mu_4)^2 \mu_4 \mu_5 \rho^2 (\mu_1 \mu_2
 +\rho^2)^2 (\mu_2 \mu_3+\rho^2)^2
 \nonumber
\\
 && \phantom{-} \times
  (\mu_2 \mu_5+\rho^2)^2
\Big(c_1^2 \mu_1^2 \mu_3 \mu_4 (\mu_1-\mu_5)^2-(\mu_1-\mu_3)^2
\mu_5
 (\mu_1 \mu_4+\rho^2)^2\Big)^2
 \nonumber
\\
  &\le& 0\;,
\eeal{Hx.delta}
the last inequality being a consequence of the non-negativity
of the $\mu_i$'s. Therefore, if a real root exists away from
the axis $\mcA$, then $\Delta=0$ at the root and $c_1^2$
satisfies there
\bel{c_1.1} c_1^2= \frac{(\mu_1-\mu_3)^2 \mu_5 (\mu_1
\mu_4+\rho^2)^2}{\mu_1^2 \mu_3 \mu_4 (\mu_1-\mu_5)^2}\;. \ee
On the other hand, the smoothness of the metric at $\rho=0$
implies (compare~(\ref{21VII.4}))
\bel{c_1.2}
 c_1^2=L^2\frac{2\kappa_1\kappa_2}{\kappa_3}\;,
\ee
where, following~\cite{EF}, $L$ is a  scale factor chosen to be
$L^2=a_2-a_1$.   We rewrite \eq{c_1.2} with the help of
\eq{21IV.43},
\bel{c_1.2s} c_1^2=\frac{(\mu_3-\mu_1 ) (\mu_4 -\mu_1 ) \mu_5
(\mu_1 \mu_3 + \ \rho^2)(\mu_1 \mu_4 + \rho^2)}{  \mu_1 \mu_3
\mu_4 (\mu_5-\mu_1 ) (\mu_1 \mu_5 + \
 \rho^2)}\;.
\ee
Subtracting \eq{c_1.1} from \eq{c_1.2s} leads to the equation
\bean
 \lefteqn{-\frac{(\mu_1 - \mu_3)
\mu_5 (\mu_1^2 + \rho^2) (\mu_1 \mu_4 + \ \rho^2) }{\ \mu_1^2
\mu_3 \mu_4 (\mu_1 - \mu_5)^2 (\mu_1 \mu_5 + \rho^2)
 }\times}&& \\
 &&\Big( \mu_1 \mu_3 (\mu_1 - \mu_4) + \mu_1 (\mu_4 -
 \mu_3) \mu_5 + (\mu_1 - \mu_3) \rho^2\Big)=0\;.
\eeal{c_1.comp}
It follows from \eq{21IV.41.1}, \eq{21IV.45}, and from
non-negativity of $\mu_i$ that each term in the last line of
\eq{c_1.comp} is strictly negative away from $\mcA$. We
conclude that this equation can only  be satisfied for
$\rho=0$, hence $H_x$ is non-zero for $\rho\neq 0$.

\subsection{Regularity for $\rho>0$}
 \label{SRegularity13VI}
  \ptclater{why does this not work in four dimensions to prove
  regularity away from the axis of the Manko solutions?}

In this section we wish to prove that the Black Saturn metrics
are regular away from the axis $\rho=0$. For this it is
convenient to review the three-soliton construction
in~\cite{EF}. The metric \eq{SaturnMetric} was obtained by a
``three-soliton transformation", a rescaling, and a
redefinition of the coordinates.%
\footnote{It has been mentioned at the end of Sec. 2.2 of
\cite{EF} that the same solution can also be obtained (in a
slightly different form) as a result of a (simpler) two soliton
transformation.}
The following generating matrix
\bea\label{psi_0}
  \Psi_0(\lambda,\rho,z)
  =\diag\left\{\frac{1}{(\mu_4-\lambda)},
    \frac{(\mu_1-\lambda)(\mu_4-\lambda)}
    {(\mu_2-\lambda)(\mu_5-\lambda)},
    -\frac{(\mu_3-\lambda)}{(\bar{\mu}_5-\lambda)}\right\}
\eea
was used, starting with the seed solution
\begin{equation}\label{G_0}
  G_0=\diag\left\{\frac{1}{\mu_4},
   \frac{\mu_1\mu_4}{\mu_2\mu_5},
   -\frac{\mu_3}{\bar{\mu}_5}\right\}\;,
\end{equation}
where $\bar\mu_5=-\rho^2/\mu_5$.
The general $n$-soliton transformation yields a
new solution $G$ with components
\begin{equation}
  G_{ab}=(G_0)_{ab}-
  \sum_{k,l=1}^{n}\frac{
  (G_0)_{ac}\, m_c^{(k)}\,  (\Gamma^{-1})_{kl}\;  m_d^{(l)}\, (G_0)_{db}}
                       {\tilde\mu_k\tilde\mu_l}
\label{eqn:unG}
\end{equation}
(the repeated indices $a,b,c,d=1,\dots,D-2$ are summed over).
The components of the vectors $m^{(k)}$ are
\begin{equation}\label{mk}
  m_a^{(k)}=m_{0b}^{(k)}\left[\Psi_0^{-1}(\tilde\mu_k,\rho,z)\right]_{ba}\, ,
\end{equation}
where $m_{0b}^{(k)}$ are the ``BZ parameters". The symmetric
matrix $\Gamma$ is defined as
\begin{equation}
  \label{gamma}
  \Gamma_{kl}=\frac{m_a^{(k)}\, (G_0)_{ab}\, m_b^{(l)}}
   {\rho^2+\tilde\mu_k \tilde\mu_l}\, ,
\end{equation}
and the inverse $\Gamma^{-1}$ of $\Gamma$ appears in
\eq{eqn:unG}. Here $\tilde \mu_i$ stands  for $\mu_i$ for those
$i$'s which correspond to solitons, or $\bar \mu_i$ for the
antisolitons, where
$$
 \bar \mu_i = -\sqrt{\rho^2+(z-a_i)^2}-(z-a_i)
 \;.
$$

The three-soliton transformation is performed in steps:

\begin{itemize}
\item Add an anti-soliton at $z=a_1$ (pole at
    $\lambda=\bar\mu_1$) with BZ vector $m_0^{(1)}
    =(1,0,c_1)$,
\item add a soliton at  $z=a_2$ (pole at $\lambda=\mu_2$)
    with BZ vector $m_0^{(2)} =(1,0,c_2)$, and
\item add an anti-soliton at $z=a_3$ (pole at
    $\lambda=\bar\mu_3$) with BZ vector $m_0^{(3)}
    =(1,0,0)$.
 \end{itemize}

Recall  the ordering $a_1<a_5<a_4<a_3<a_2$, and we impose the
regularity condition \eq{c_1.2}. Using these assumptions, we
show that that the procedure described above leads to a smooth
Lorentzian metric on $\{\rho>0\}$.

Firstly, we note that
\begin{itemize}
\item $\mu_i-\mu_k\neq 0$ for $i\neq k$ and $\rho>0$,
\item $\mu_i - \bar\mu_k\neq 0$ for $\rho>0$,
\end{itemize}
where the first point follows from \eq{21IV.42}. The second
statement is a consequence of:
$\mu_i-\bar\mu_k=\sqrt{\rho^2+(z-a_i)^2}+\sqrt{\rho^2+(z-a_k)^2}+a_i-a_k$,
hence $\mu_i-\bar\mu_k=0$ implies
$(a_i-a_k)^2=(\sqrt{\rho^2+(z-a_i)^2}+\sqrt{\rho^2+(z-a_k)^2})^2$,
which is equivalent to
$$
 \rho^2+\sqrt{\rho^2+(z-a_i)^2}\sqrt{\rho^2+(z-a_k)^2}+(z-a_i)(z-a_k)=0
 \;.
$$
The middle term dominates the absolute value of the last one,
which implies that the last equality is satisfied if and only
if $\rho=0$ and $(z-a_i)(z-a_k)\le 0$, in particular it cannot
hold for $\rho>0$.

We conclude that $\psi_0^{-1}$ is analytic in $\rho$ and $z$ on
$\{\rho >0\big\}$.
Subsequently the components of the vectors $m^k$ are analytic
there (see \eq{mk}) and so is the matrix $\Gamma$ (see
\eq{gamma}). The n-soliton transformation \eq{eqn:unG} contains
$\Gamma^{-1}$, thus $\det\Gamma$ appears in denominator in all
terms in sum in \eq{eqn:unG} (excluding $(G_0)_{ab}$). Since
the numerator of these terms contains analytic expressions and
a cofactor of $\Gamma$, then only the vanishing of $\det\Gamma$
may lead to singularities in the metric coefficients $g_{ab}$
on $\{\rho >0\big\}$. We show below that $\det\Gamma$ does not
have zeros there provided that the free parameters satisfy the
regularity conditions \eq{c_1.2}. This will prove that the
metric functions $g_{tt}$, $g_{t\psi}$ and $g_{\psi\psi}$ are
smooth away from $\{\rho=0\}$. Hence
$$
 \frac{H_y}{H_x}\;,
 \quad
 \frac{\omega_\psi}{H_x}\;,
 \quad
 \frac{H_y}{H_x}\left( \big(\frac{\omega_\psi}{H_y} +q\big)^2 -\frac{G_y H_x}{H_y}\right)\;,
$$
are smooth for $\rho>0$. This is equivalent to smoothness, away
from the axis, of the set of functions
$$
 \frac{H_y}{H_x}\;,
 \quad
 \frac{\omega_\psi}{H_x}\;,
 \quad
 \frac{{\omega_\psi^2} }{H_yH_x}
 \;.
$$
Since $H_x$ has been shown to have no zeros away from the axis,
we also conclude that
$$
 \frac{\omega_\psi^2}{H_y}
 $$
is smooth away from $\rho=0$.

The next steps in the construction of the line element
\eq{SaturnMetric} involve a rescaling by
$\rho^2\frac{\mu_2}{\mu_1\mu_3}$ and a change of $t$, $\Psi$
coordinates $t\rightarrow t-q\Psi$, $\Psi\rightarrow-\Psi$.
These operations do not affect the regularity of the metric
functions.

Let us now pass to the analysis of $\det \Gamma$. The metric
functions $g_{\rho\rho}=g_{zz}$, denoted as $e^{2\nu}$
in~\cite{EF},
can be calculated using a formula of Pomeransky
\cite{Pomeransky}:
\begin{equation}
H_x k^2 P \equiv e^{2\nu}=e^{2\nu_0}\frac{\det\Gamma}{\det\Gamma_{kl}^{(0)}},
\end{equation}
where~\cite{EF}
\bea
   e^{2\nu_0}=  k^2\,
   \frac{\mu_2 \, \mu_5
        (\rho^2+\mu_1\,\mu_2)^2(\rho^2+\mu_1\,\mu_4)
        (\rho^2+\mu_1\,\mu_5)(\rho^2+\mu_2\mu_3)
        (\rho^2+\mu_3\,\mu_4)^2 (\rho^2+\mu_4\,\mu_5)}
     {\mu_1 (\rho^2+\mu_3\mu_5)(\rho^2+\mu_1\,\mu_3)
      (\rho^2+\mu_2\,\mu_4)(\rho^2+\mu_2\,\mu_5)
     \prod_{i=1}^{5}(\rho^2+\mu_i^2)}\, ,
     \nonumber\\\label{Hexp2nuB}
\eea
and where $\Gamma^{(0)}$ corresponds to $\Gamma$ with
$c_1=c_2=0$. But from what has been said the functions
$\det\Gamma^{(0)}$ and $P$ do not have zeros for $\rho>0$.
Since we have shown that $H_x$ does not have zeros there, the
non-vanishing of $\det \Gamma$ follows.

We conclude that the metric functions appearing in the Black
Saturn metric \eq{SaturnMetric} are analytic for $\rho>0$. It
remains to check that the resulting matrix has Lorentzian
signature. This is clear at large distances by the asymptotic
analysis of the metric in Section~\ref{ssAatinf13VII}, so the
signature will have the right value if and only if the
determinant of the metric has no zeros. This determinant equals
\bel{2XII.2}
 \det g_{\mu\nu} = -\rho^2 H_x^2 k^4 P^2
 \;.
\ee
and its non-vanishing for $\rho>0$ follows from
Section~\ref{ssPHx}.

\subsection{The ``axis" $\{\rho=0\}$}
 \label{ssAor13VII}

The regularity of the metric functions on the axis $\{\rho=0\}$
requires separate attention. The behaviour, near that axis, of
the functions that determine the metric depends strongly on the
part of the $z$ axis which is approached. For example, the
$\mu_i$'s are identically zero for $z\ge a_i$ at $\rho=0$, but
are not  for $z<a_i$. This results in an intricate behaviour of
the functions involved, as illustrated by Tables~\ref{T18VII.1}
and \ref{T18VII.0}.

\begin{table}
\begin{center}
\begin{equation}\nonumber
\hspace{-1cm}
\begin{array}{||c|c|c|c||}
\hline
\hline
 z & P
        & G_x= g_{\varphi\varphi}     = \frac{\rho^2 \mu_4}{\mu_3\mu_5}
        & G_y       = \frac{\mu_3\mu_5} {  \mu_4}
\\
\hline
\hline
 z <a_1 & 2^8(z-a_3)^2(z-a_4)^3(z-a_1)(z-a_5)^2
        &                         -\frac{z-a_4} {2(z-a_3)(z-a_5)}\rho^2
        &                         -\frac{2(z-a_3)(z-a_5)} {z-a_4}
\\
\hline
 a_1<z < a_5 & 2^6(z-a_3)^2(z-a_4)^3(z-a_5)\left(\frac{a_5-a_1}{z-a_1}\right) \rho^2
        &                         -\frac{z-a_4} {2(z-a_3)(z-a_5)}\rho^2
        &                         -\frac{2(z-a_3)(z-a_5)} {z-a_4}
\\
\hline
a_5<z < a_4 & 2^4(z-a_3)^2(z-a_4)^2\left(\frac{a_4-a_5}{z-a_5}\right) \rho^4
        &                          \frac{2(z-a_4)(z-a_5)}{z-a_3}
        &                          \frac{z-a_3}{2(z-a_4)(z-a_5)} \rho^2
\\
\hline
 a_4<z < a_3 & \left(\frac{a_3-a_4}{z-a_4}\right)^2 \rho^8
        &                          -\frac{z-a_5}{2(z-a_4)(z-a_3)}   \rho^2
        &                          -\frac{2(z-a_4)(z-a_3)}{z-a_5}
\\
\hline
 a_3 < z < a_2 & \rho^8
        &                          \frac{2(z-a_3)(z-a_5)}{z-a_4}
        &                          \frac{z-a_4}{2(z-a_3)(z-a_5)} \rho^2
\\
\hline
a_2< z & \rho^8
        &                          \frac{2(z-a_3)(z-a_5)}{z-a_4}
        &                          \frac{z-a_4}{2(z-a_3)(z-a_5)}   \rho^2
\\
\hline                             \hline
\end{array}
\end{equation}
\end{center}
\caption{\label{T18VII.1}%
 Leading order behaviour near $\rho=0$ of $P$, $G_x$ and $G_y$.}
\end{table}
\begin{table}
\begin{center}
 \begin{equation}\nonumber
 \hspace{-2cm}
 \begin{array}{||c|c|c|c||}
\hline
\hline
 z & H_x
        & g_{\varphi \varphi}/g_{\rho \rho} = \frac{G_x}{H_x k^2 P }
\\
\hline
\hline
 z < a_1
 &      -\frac{(2 (a_{1}-a_{3}) (a_{2}-a_{4})+(a_{1}-a_{5}) c_{1} c_{2})^2}{2^{11} (a_{1}-a_{3})^2 (a_{2}-a_{4})^2 (a_{1}-z) (a_{3}-z)^3 (a_{4}-z)^2 (z -a_{5})^3}
 &       \frac{2^{2} (a_{1}-a_{3})^2 (a_{2}-a_{4})^2  }
    {(2 (a_{1}-a_{3}) (a_{2}-a_{4})+(a_{1}-a_{5}) c_{1} c_{2})^2k^2}   \rho^2=\rho^2
\\
\hline
 a_1<z < a_5 &    \frac{(a_2 c_1-a_1 c_2+a_4 (c_2 -c_1))^2 (z-a_1)}{2^8  (a_1-a_3) (a_1-a_4) (a_2-a_4)^2 (a_3-z)^
 3 (a_4-z)^2 (a_5-z)^2 } \rho^{-2}
        &                   \frac{2 (a_3-a_1) (a_4-a_1)(a_4-a_2)^2}{(a_5-a_1) (a_2 c_1-a_1 c_2+a_4 (c_2-c_1))^2k^2} \rho^{2}=\rho^2
\\
\hline
a_5<z < a_4 & \sim\rho^{-4}
        &              \sim 1           \quad      \text{(black ring horizon ?)}
\\
\hline
 a_4<z < a_3 & \frac{(a_1-a_2)^2(a_4-z)(z-a_5)}{2(a_1-a_3)(a_2-a_4)(a_2-a_5)(a_3-a_5)(a_3-z)}\rho^{-8}
        & -\frac{(a_1-a_3)(a_2-a_4)(a_2-a_5)(a_3-a_5)}{(a_1-a_2)^2(a_3-a_4)^2k^2}\rho^2	=\rho^2
\\
\hline
 a_3 < z < a_2 & \sim\rho^{-8}
        &              \sim 1     \quad \text{(spherical horizon ?)}
\\
\hline
a_2 < z
        & \sim\rho^{-8}
        & \sim 1
\\
\hline                             \hline
\end{array}
\end{equation}
\caption{\label{T18VII.0}%
Leading order behaviour near $\rho=0$ of $H_x$ and of
$g_{\varphi \varphi}/g_{\rho\rho}$. The value $1$ of the
coefficient in front of $\rho^2$ is precisely what is needed
for absence of conical singularities at the axis. We write
$f\sim \rho^\alpha$, for some $\alpha\in \R$, if the leading
order behaviour of $f$, for small $\rho$, is $f=C\rho^\alpha$,
for some constant $C$ depending upon the parameters at hand,
the exact form of which was too long to be displayed here. The
question marks concerning the horizons are taken care of in
Section~\ref{ssHor13VII}--\ref{ss24IV10.1}.}
\end{center}
\end{table}
%
 \ptc{last sentence added to the caption of the second table}

\subsubsection{$g_{\varphi\varphi}$} A complete description of
the behaviour of $g_{\varphi\varphi}$ at $\rho=0$ can be found
in Table~\ref{T18VII.1}. One can further see from
Table~\ref{T18VII.0} that the Killing vector field
$\partial_\varphi$ has a smooth axis of rotation on $\{\rho=0,
z< a_5\}\cup \{\rho=0, a_4<z<a_3\}$, as already discussed in
Section~\ref{s6XII.1}.

\subsubsection{$g_{tt}$}
At $\rho=0, z<a_1$, the metric function $g_{tt}$ is a rational
function of $z$ with denominator

\bel{6XII.1}
 \alpha (a_1 - z) (a_2 - z) (a_4 - z)
 \;,
\ee
where
%
\beaa
 \alpha
 &:=&(2 (a_1-a_3) (a_2-a_4)+c_1 c_2 (a_1-a_5))^2
\\
 &=&
  \frac{4 (a_1-a_2)^2 (a_1-a_3)
(a_2-a_4)
(a_3-a_4)^2}{(a_2-a_5)
(a_5-a_3)}
 \;.
\eeaa
So $\alpha$ is nonzero when all the $a_i$'s are distinct. We
have already seen that the singularity at $z=a_1$ is removable;
the ones suggested by \eq{6XII.1} at $a_2$ and $a_4$ are
irrelevant at this stage, since we have assumed $z<a_1$ to
obtain the expression.

From what has been proved in Section~\ref{Schoicec_1}, $g_{tt}$
extends analytically across $z=a_1$, so the last analysis
applies on $\rho=0, a_1<z<a_5$.

The zeros of the  {denominator} of $g_{tt}$ restricted to
$\rho=0$, $a_5<z<a_4$ turn out not to be obvious. It should be
clear from the form of $g_{tt}$ that those arise from the zeros
of the numerator of $H_x$. {This numerator turns out to be a
complicated polynomial in the
$a_i$'s, $z$, and the $c_i$'s, quadratic in $c_2$.}%
\footnote{The reader is warned that the numerators listed below
depend upon whether or not the constants $c_a$ and $k$ have
been replaced by their values in terms of the $a_i$'s.}
As in Section~\ref{Schoicec_1}, we calculate the discriminant
of this polynomial, which reads
%
%
 $$
 8 (a_1-a_2)^2 (a_1-a_4)^4
 (a_1-z)^2 (a_2-a_4)^2 (a_2-a_5)^2
 (a_3-z) (a_4-z) (a_5-z)
 \;,
 $$
and which is negative because of the last factor. We conclude
that $g_{tt}$ does not have poles in $(a_5,a_4)$.

 The apparent pole at $z=a_5$  above is removable:  Indeed one can compute the limit $z \to
a_5^-$ using the formula for $g_{tt}$ at $\rho=0$, $z \in
(a_1,a_5)$. After $c_1$ is substituted, one obtains a rational
expression with denominator
\begin{align}
 (a_2-a_5) (a_1-a_5) (a_4-a_5)
 \left(\sqrt{\frac{(2(a_1 - a_3) (a_4 - a_1))}{(a_1 - a_5)}} (a_2-a_4) + (a_4-a_1) c_2\right)^2
 \;.
\end{align}
Substituting $c_2$ into the expression above we obtain
$$\frac{2 (a_1 -a_2  )^2 (a_1 -a_4  ) (a_2  -a_4  ) (a_3  -a_4  )^2 (a_4  -a_5 )
}{a_3  -a_5 }
 \;,
$$
which  does not vanish provided that all the $a_i$'s are
different. The same value of $g_{tt}$ is obtained by taking the
limit $z \to a_5^+$ for $g_{tt}$ in region $\rho=0$, $z \in
(a_5,a_4)$. So we conclude that $g_{tt}|_{\rho=0}$ is
continuous at  $z = a_5$. A similar calculation establishes
continuity of $g_{tt}|_{\rho=0}$ at  $z = a_4$; here the
relevant denominator of the limit $z \to a_4^-$ reads:
%
$$
2 (a_2 - a_1)^2 (a_2 - a_4) (a_4 - a_1) (a_4 - a_5)
\;.
$$
%

%
The  {denominator} of $g_{tt}$ restricted to $\rho=0$,
$a_4<z<a_3$ can be written as
%
$$
2 ({a_1}-{a_2})^2 (a_1-z) ({a_2}-z) (z-{a_5})
 \;,
$$
and is therefore smooth on this interval, extending
continuously to the end points.

Non-existence of zeros of the denominator of $g_{tt}$
restricted to $\rho=0$, $a_3<z<a_2$ can be proved similarly as
for $a_5<z<a_4$.  After factorisations and cancellations, the
numerator of $H_x$ there is
a complicated polynomial in the $a_i$'s, $z$, and the
$c_i$'s, quadratic in $c_2$. The discriminant of this
polynomial equals
 $$
8 ({a_1}-a_2)^2 ({a_1}-{a_3})^4 ({a_1}-z)^2 ({a_2}-{a_3})^2 ({a_2}-{a_5})^2 ({a_3}-z) ({a_4}-z) ({a_5}-z)
 \;,
 $$
which is negative because of the third-to-last factor. We
conclude that $g_{tt}$ is smooth in a neighbourhood of
$\{\rho=0,z\in(a_3,a_2)\}$. The continuity of $g_{tt}|_{\rho
=0}$ at $z = a_3$ may again be checked by taking left and right
limits.

Non-existence of zeros of the denominator of $g_{tt}$
restricted to $\rho=0$, $a_2<z$ can again be proved by
calculating a discriminant. The numerator of $H_x$ there is
a quadratic polynomial in   $c_2$, with discriminant
 $$
32 (a_1-{a_2})^2 (a_1-{a_3})^4 ({a_1}-z)^2 ({a_2}-{a_4})^2
 ({a_3}-z) ({a_4}-z) ({a_5}-z) \;.
 $$
This is negative because each of the three last factors is
negative. We conclude that $g_{tt}$ is smooth on a
neighbourhood of $\{\rho=0, z\in (a_2,\infty)\}$.

\subsubsection{Ergosurfaces}
 \label{ssErgo25III.1}

The \emph{ergosurfaces} are defined as the boundaries of the
set $g_{tt}\le  0$. Their intersections with the axis are
therefore determined by the set where $g_{tt}$ vanishes on the
axis. We will not undertake a systematic study of those, but
only make some general comments; see~\cite{EFHHR} for some
results concerning this issue.

Near the points $a_i$  the numerator of $g_{tt}$ has the
following behaviour:
\begin{center}
$\sim c_2^2$ for $a_1$ (see \eqref{21VII.1}),\\
\vspace{0.3cm}
$\sim((a_2  -a_4  ) (a_1 -a_5 ) c_1+( a_4 -a_1  ) (a_2  -a_5 ) c_2)^2$ for $a_5$, $a_4$,\\
\vspace{0.3cm}
$\sim((a_2  -a_3  ) (a_1 -a_5 ) c_1+( a_3 -a_1  ) (a_2  -a_5 ) c_2)^2$ for $a_3$,\\
\vspace{0.3cm}
$\sim((a_2  -a_3  ) (a_1 -a_5 ) c_1+( a_3 -a_1  ) (a_2  -a_5 ) c_2)^2 (a_2  -z)$ near $a_2^-$,\\
\vspace{0.3cm} $\sim(2 (a_1 -a_3  ) (a_2  -a_4  )+(a_1 -a_5 )
c_1 c_2)^2(a_2  -z)$ near $a_2^+$,
\end{center}
where $\sim$ stands for a manifestly non-vanishing
proportionality factor.
This shows that a component of the ergosurface always
intersects the axis at $z=a_2$. It also  follows from  the
above that the intersection of the ergosurface with the axis
$\{\rho=0\}$ contains $z=a_1$ and $z=a_2$ as isolated points
when $c_2=0$.

Next, a {\sc Mathematica} calculation (in which $c_1$ has been
replaced by its values in terms of the $a_i$'s) shows that on
$(-\infty,a_5)$ the metric function $g_{tt}|_{\rho =0}$ can be
written as a rational function with numerator which is
quadratic in $z$. Recall that the numerator does not change
sign on $(-\infty,a_5)$, so $g_{tt}|_{\rho =0}$  is continuous
with at most two zeros there. But $g_{tt}|_{\rho =0}$  is
negative for large negative $z$,  while at $z=a_5$ we have
\bel{21IV10.3}
 g_{tt}(\rho=0, z=a_5) = \frac{(a_5-a_3) \left(c_1
 (a_1-a_5) (a_2-a_4)+c_2 (a_4-a_1) (a_2-a_5)\right)^2}
 {(a_5-a_1)(a_2-a_5) (a_5-a_4) \left(a_2 c_1 -a_1 c_2 +a_4
 (c_2-c_1)\right)^2}
 \;,
\ee
which is strictly positive.
We conclude
that $g_{tt}|_{\rho =0}$ always has precisely one zero on
$(-\infty,a_5)$.

\begin{figure}
\begin{center}
\includegraphics[scale=.4]{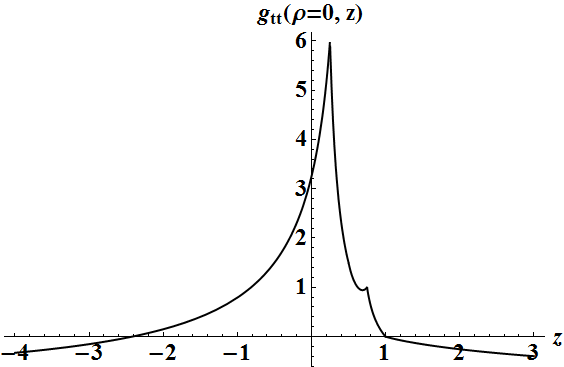}
\end{center}
 \caption{\label{F21IV10.1}$g_{tt}|_{\rho =0}$ as
a function of $z$ for $a_1=0$, $a_2=1$, $a_3=3/4$, $a_4=1/2$, $a_5=1/4$. In this case
the ergosurface encloses both horizons.
}
\end{figure}

In Figure~\ref{F21IV10.1} we show the graph of  $g_{tt}|_{\rho
=0}$ for a set of simple values of parameters.

\subsubsection{$g_{\rho\rho}$ and $g_{zz}$}
 \label{S27IV10.44}

The metric functions $g_{\rho\rho}=g_{zz}$ on $\rho=0$, $z\in
(a_1,a_5)$ equal
%
%
\bel{3I.1}
 -\frac{a_4  -z}{2 (a_3  -z) (z-a_5 )}
 \;,
\ee
and are therefore smooth there. By analyticity, the same
expression is valid for $z\in (-\infty,a_5)$.

The metric function $g_{\rho\rho}$ on $\rho=0$, $z\in
(a_5,a_4)$ can be written as a rational function of $z$, with
denominator
$$
 {4 (a_1 -a_2 )^2
(a_2 -a_4  ) (a_2 -z) (a_3  -a_4  )^2
(a_4  -z) (z-a_5 )}
 \;,
$$
%
and is thus smooth near $\{\rho=0,z\in (a_5,a_4)\}$.%
\footnote{This denominator has been obtained by substituting
the values of $k$ and $c_1$, but not $c_2$.\label{fnnoc_2sub}}
One checks that for $z>a_5$ and
close to $a_5$ we have
%
\bel{3I.2}
 g_{\rho\rho}|_{\rho=0}=\frac{a_4  -a_5 }{2
 (a_3  - a_5 )(z-a_5)} +O(1)
 \;,
\ee
leading to a pole of order one when $a_5$ is approached from
above. Comparing with \eq{3I.1} one finds that $|z-a_5| \times
g_{\rho\rho}|_{\rho=0}$ is continuous at $a_5$.

Next, for  $z<a_4$ and close to $a_4$ we have
%
\bel{3I.3}
 g_{\rho\rho}|_{\rho=0}=\frac{a_5 -a_4  }{2
 (a_3  - a_4  )(z-a_4)} +O(1)
 \;,
\ee
leading to a pole of order one when $a_4$ is approached from
 below.
%
%
%

The metric function $g_{\rho\rho}$ on $\rho=0$, $z\in
(a_4,a_3)$ equals
%
\bel{6I.1}
 -\frac{z-a_5 }{2(z-a_3  )(z-a_4  )}
\ee
%
%
with simple poles at $a_4$ and $a_3$. Comparing with \eq{3I.3}
one finds that
$$
 |z-a_4| \times g_{\rho\rho}|_{\rho=0}
$$
is continuous at $a_4$.

The metric function $g_{\rho\rho}$ on $\rho=0$, $z\in
(a_3,a_2)$ can be written as a rational function of $z$, with
denominator
$$
4 (a_1 -a_2  )^2 (a_1 -a_3  ) (a_2  -a_3  ) (a_3  -a_4  )^2 (a_1 -z) (a_2  -z) (a_3  -z) (z -a_5)
 \;,
$$
which has been obtained by substituting in $k$, but neither
$c_1$ nor $c_2$.
%
For  $z>a_3$ and close to $a_3$ we have
%
\bel{a_3  pole}
 g_{\rho\rho}|_{\rho=0}=\frac{a_3  -a_5 }{2
 (a_3  - a_4  )(z-a_3)} +O(1)
 \;,
\ee
and there is a first order pole when $z=a_3$ is approached from
above. Comparing with \eq{6I.1} one finds that $
 |z-a_3|\times  g_{\rho\rho}|_{\rho=0}
$ is continuous at $a_3$.

%
%
Again, for $z<a_2$ and close to $a_2$ we have
\bel{a_2  pole}
 g_{\rho\rho}|_{\rho=0}=
\frac{(a_1 -a_3  ) (a_3  -a_5 ) \big(2 (a_2  -a_3  ) (a_2  -a_4
)+(a_2  -a_5 ) c_2^2\big)}{4 (a_1 -a_2  ) (a_2  -a_3  ) (a_3
-a_4 )^2 (a_2  -z)} +O(1)
 \;,
\ee
Since $c_2$ is real, the numerator of the leading term does not
vanish. Therefore, $g_{\rho\rho}|_{\rho=0}$ has a first order
pole when $z=a_2$ is approached from below.

The metric function $g_{\rho\rho}$ on $\rho=0$, $z\in
(a_2,\infty)$ can be written as a rational function of $z$,
with denominator$^{\mbox{\scriptsize\ref{fnnoc_2sub}}}$
$$
4 (a_1  - a_2  )^2 (a_3   - a_4  )^2 (-a_2   + a_4  ) (a_2   - z) (a_3   - z) (-a_5  + z)\;.
$$
%
Finally, for $z>a_2$ and close to $a_2$ we have
\bel{a_2  poleplus}
 g_{\rho\rho}|_{\rho=0}=
-\frac{(a_1 -a_3  ) (a_3  -a_5 ) (2 (a_2  -a_3  ) (a_2  -a_4
)+(a_2  -a_5 ) c_2^2)}{4 (a_1 -a_2  ) (a_2  -a_3  ) (a_3  -a_4
)^2 (a_2  -z)} +O(1)
 \;.
\ee
This coincides with \eq{a_2  pole} except for an overall sign.
Again, with $c_2$ being real the numerator of the leading term
cannot vanish, so the limits from above and from below of
$|z-a_2|\times g_{\rho\rho}|_{\rho=0}$  at $z=a_2$ are
different from zero, and coincide.

\subsubsection{$g_{t\psi}$ and $g_{\psi\psi}$}
%
%
%

We pass now to the singularities of
$$g_{t \psi}= -\frac{H_y}{H_x}\left(\frac {\omega_\psi}{H_y}+q\right)
$$
on the axis $\rho=0$. It  turns out that  the calculations here
are very similar to those for $g_{tt}$, keeping in mind that
the interval $(-\infty,a_5 )$ was handled in Section
\ref{Schoicec_1}. In particular the lack of zeros of the
relevant denominators on each subinterval of the $z$--axis is
established in exactly the same way as for  $g_{tt}$, while
continuity at the $a_i$'s is obtained by checking the left and
right limits.
%
%
This results most likely from the rewriting
$$
 g_{t \psi}= - \frac {F\omega_\psi +q F H_y}{FH_x}
 \;,
$$
and noting that, away from the $a_i$'s,  any infinities   of
$g_{t\psi}|_{\rho=0}$ can only result from zeros of $FH_x$. In any
case, a {\sc Mathematica} calculation shows that no further
infinities in $g_{t \psi}|_{\rho=0}$ arise on the axis from
$F\omega_\psi +q F H_y$, and in fact the denominators of $g_{t
\psi}|_{\rho=0}$, when this last function is written as a
rational function of the $z$'s, $a_i$'s, and the $c_i$'s,
\emph{coincide} with those of $g_{tt}|_{\rho=0}$.
So, we find that   $g_{t \psi} $ is smooth near
\bel{7I.1} I:=\{\rho=0, z\in (-\infty,a_5 )\cup(a_5 ,a_4
)\cup(a_4 ,a_3  )\cup(a_3  ,a_2  )\cup(a_2  ,+\infty)\}
 \;.
\ee
%
%

For the remaining points $a_2,\ldots,a_5$, we write instead
\bel{25III10.1}
 g_{t\psi}=g_{tt}\left(\frac{\omega_\psi}{H_y}+q\right)
 \;.
\ee
Using {\sc Mathematica} we verified that the left and right
limits of $(\omega_\psi/H_y)|_{\rho =0}$ at $a_{i=1,5,4,3}$ are
equal, but the left and right limit at $a_2$ is not. These are,
respectively:
\begin{center}
$\frac{2 (a_2   - a_4  )}{c_2}$ for $a_1$,\\
\vspace{0.3cm}
$\frac{2 (a_1  - a_2  ) (a_1  - a_4  ) (a_2   - a_4  )}{(a_2   - a_4  ) (a_1  - a_5 ) c_1 + ( a_4 -a_1   ) (a_2   - a_5 ) c_2}$ for $a_5$, $a_4$,\\
\vspace{0.3cm}
$\frac{2 (a_1  - a_2  ) (a_1  - a_3  ) (a_2   - a_3  )}{(a_2   - a_3  ) (a_1  - a_5 ) c_1 + ( a_3-a_1  ) (a_2   - a_5 ) c_2}$ for $a_3$, $a_2^-$,\\
\vspace{0.3cm}
$\frac{2 (a_1 -a_2  ) (a_1 -a_3  ) c_2}{2 (a_1 -a_3  ) (a_2  -a_4  )+(a_1 -a_5 ) c_1 c_2}$ for $a_2^+$.\\
\end{center}
(Note that the first line above contains an inverse power of
$c_2$, and so the case $c_2=0$ requires separate attention;
this is handled in Section~\ref{ssAor25III10}). {On the other
hand, the numerator of $g_{tt}$ on $\rho=0$ has already been
analysed in Section~\ref{ssErgo25III.1}, we repeat the formulae
for the convenience of the reader
\begin{center}
$\sim c_2^2$ for $a_1$ (see \eqref{21VII.1}),\\
\vspace{0.3cm}
$\sim((a_2  -a_4  ) (a_1 -a_5 ) c_1+( a_4-a_1  ) (a_2  -a_5 ) c_2)^2$ for $a_5$, $a_4$,\\
\vspace{0.3cm}
$\sim((a_2  -a_3  ) (a_1 -a_5 ) c_1+( a_3-a_1  ) (a_2  -a_5 ) c_2)^2$ for $a_3$,\\
\vspace{0.3cm}
$\sim((a_2  -a_3  ) (a_1 -a_5 ) c_1+( a_3-a_1  ) (a_2  -a_5 ) c_2)^2 (a_2  -z)$ near $a_2^-$,\\
\vspace{0.3cm} $\sim(2 (a_1 -a_3  ) (a_2  -a_4  )+(a_1 -a_5 )
c_1 c_2)^2(a_2  -z)$ near $a_2^+$.
\end{center}
We note that the $z$-independent terms above  all have the same
sign when $c_1c_2>0$, hence they are not identically zero. Thus
the factors displayed here in the numerator of $g_{tt}$
can be cancelled with the corresponding factors in the denominator
in the product $g_{tt} \times
(\omega_\psi/H_y)$ arising in \eqref{25III10.1}.
%
This implies that
$g_{t\psi}|_{\rho=0}$ is continuous for $z\in\R $.

Consider next $g_{\psi\psi}|_{\rho=0}$,
$$
 g_{\psi\psi}=g_{tt}\left(\frac {\omega_\psi}{H_y}+q\right)^2-\frac{G_y}{g_{tt}}
 \;.
$$
A {\sc Mathematica} calculation shows again that the
denominator of this function, when written as a rational
function of $z$ and the $a_i$'s, coincides with the denominator
of $g_{tt}|_{\rho=0}$, which has already been shown to have no
zeros. This, implies that $g_{\psi\psi}|_{\rho=0}$ is smooth
near the set appearing in \eq{7I.1}.

\begin{figure}
\begin{center}
\includegraphics[scale=.4]{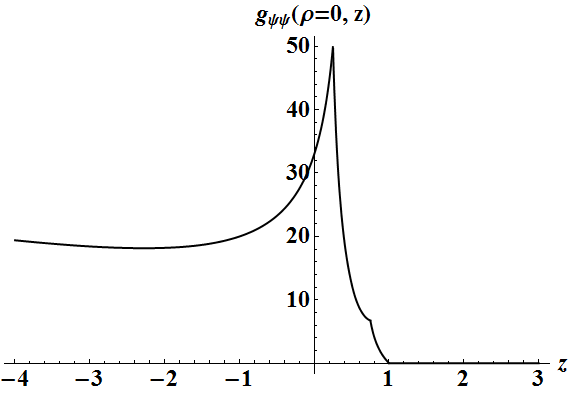}
\end{center}
 \caption{\label{FgPP} The graph of $g_{\psi\psi} $ on the axis for $a_1 =0, a_5=\tfrac{1}{4}, a_4=\tfrac{1}{2}, a_3=\tfrac{3}{4}, a_2=1.$}
\end{figure}

From what has been said so far, to prove continuity of
$g_{\psi\psi}$ it remains to establish continuity of
$G_y/g_{tt}$ at $z=a_i$.} Now, $G_y$ is continuous on $\rho=0$
for $z\in\R $ and vanishes for $z\geq a_3$ (see
Table~\ref{T18VII.1})
%
so $g_{\psi\psi}|_{\rho=0}$ is continuous at $
\{a_5,a_4,a_3,a_2\}$. We conclude that $g_{\psi\psi}$ is smooth
near the set in \eq{7I.1}, and that $g_{\psi\psi}|_{\rho=0}$ is
continuous at all $z\in\R$.

However, the above is not the whole story about $g_{\psi\psi}$,
as we need to know where $g_{\psi\psi}|_{\rho=0}$ vanishes;
such points correspond  either to lower dimensional orbits, or
to closed null curves.

It already follows implicitly from Section~\ref{ssAatinf13VII}
that $g_{\psi\psi}|_{\rho=0}=0$ for $z>a_2$ and, in fact, in
that interval of $z$'s we have
\bel{24III10.1}
 g_{\psi\psi} = g_{\rho \rho}(1+O(\rho^2))\rho^2
 \;,
\ee
as needed for a regular ``axis of rotation". This formula is
obtained by a direct {\sc Mathematica} calculation, in the
spirit of the ones already done in this section. We emphasize
that we are not claiming uniformity of the error term
$O(\rho^2)$ above as $a_2$ is approached.

Note that $g_{\rho\rho}>0$ away from the axis, and it follows
from \eq{24III10.1} that $g_{\psi\psi}>0$ for $z>a_2$ and $\rho
>0$ small enough.

The question of the sign of $g_{\psi\psi}|_{\rho=0}$   on the
remaining axis intervals is addressed in
Section~\ref{ss16IV10.2} under the hypothesis that $c_2=0$. In
Appendix~\ref{nr} we give numerical evidence that
$g_{\psi\psi}|_{\rho=0}$ is positive on $\{z< a_2\}$ for
general $c_2$'s, see Figure~\ref{fig:plot1S}. The values of
$g_{\psi\psi}|_{\rho=0}$ at $z=a_i$ for $i=5,4,3$ can be easily
obtained by direct limits computation. As expected from the
continuity established earlier the right and left limits
coincide and are equal to
\begin{center}
$\frac{(a_5-a_3) (q (c_1 (a_1-a_5) (a_2-a_4)+c_2 (a_4-a_1) (a_2-a_5))+2 (a_1-a_2)
   (a_1-a_4) (a_2-a_4))^2}{(a_1-a_5) (a_5-a_2) (a_5-a_4) (a_2
   c_1 -a_1 c_2 +a_4 (c_2-c_1))^2}$ for $a_5$,\\
\vspace{0.3cm}
$\frac{(q (c_1 (a_1-a_5) (a_2-a_4)+c_2 (a_4-a_1) (a_2-a_5))+2 (a_1-a_2) (a_1-a_4)
   (a_2-a_4))^2}{2 (a_1-a_2)^2 (a_1-a_4) (a_4-a_2) (a_4-a_5)}$ for $a_4$,\\
\vspace{0.3cm}
$\frac{(q (c_1 (a_1-a_5) (a_2-a_3)+c_2 (a_3-a_1) (a_2-a_5))+2 (a_1-a_2) (a_1-a_3)
   (a_2-a_3))^2}{2 (a_1-a_2)^2 (a_1-a_3) (a_3-a_2) (a_3-a_5)}$ for $a_3$.
\end{center}
From the ordering of $a_i$'s \eq{21IV.41.1} it follows  that
$g_{\psi\psi}(\rho=0,z=a_i) > 0$ for $i=5,4,3$ if the
parameters are distinct. \ptclater{Niestety nie wiadomo, co sie
dzieje pomiedzy...}

Finally, we need to check the signature of the metric. A {\sc
Mathematica} calculation shows that near $I$, as defined in
\eq{7I.1}, we can write
\bel{7I.3}
 \det g_{\mu\nu} = (f +O(\rho^2))\rho^2
 \;,
\ee
where $f$ is an analytic function of $z$;
for example,
\bel{15IV10.8}
 f=\left\{
     \begin{array}{ll}
       \frac{z -a_4 }{2 (a_3  -z) (z -a_5)}\;, & \hbox{$z<a_5$\;;} \\
       \frac{z -a_5 }{2 (a_3  -z) (z -a_4)}\;, & \hbox{$a_4<z<a_3$\;.}
     \end{array}
   \right.
\ee
(No uniformity near the end points is claimed for the error
term in \eq{7I.3}.) The explicit formulae for $f$ on the
remaining intervals are too long to be usefully cited here. We
simply note that we already know that the determinant of the
metric  is strictly negative for $\rho>0$, and thus $f\le 0$ on
the axis by continuity. However, $f$ could have zeros, which
need to be excluded. Clearly there are no such zeros in the
intervals listed in \eq{15IV10.8}. Next, in the region $z
> a_2$ one finds that $f=-h^2$, where $h$ is a
quadratic function of $c_2$. The discriminant of $h$ with
respect to $c_2$ reads
$$
32 (a_1 - a_2)^2 (a_1 - a_3)^4 (a_2 - a_4)^2 (a_1 - z)^2 (a_3 - z) (a_4 - z) (a_5 - z)
 \;.
$$
This is strictly negative for $z > a_2$ and we conclude that
$f$ does not vanish on this interval.

Taking into account the polar character of the coordinates
$(\rho,\varphi)$ and $(\rho, \psi)$ near the relevant intervals
of $z$, what has been said so far together with formula
\eq{7I.3} implies that $g$ is a smooth Lorentzian metric on
$$
 \R^4\setminus\{\rho =0\;, \ z \in [a_5,a_4]  \cup [a_3,a_2]\}
 \;.
$$
The missing open intervals, and their end points, need separate
attention; this will be addressed in Sections~\ref{ssHor13VII}
and \ref{ssIntAxHor13VII}.

\subsection{Extensions across Killing horizons}
 \label{ssHor13VII}

It is expected that the interval $z\in[a_5,a_4]$  lying on the
coordinate axis $\rho=0$,  corresponds to a ring Killing
horizon with  topology $\R\times S^1\times S^2$, while
$z\in[a_3,a_2]$ corresponds to a spherical Killing horizon,
with  topology $\R\times S^3$. The aim of this section is to
establish this, modulo possibly the end points where the axis
meets the Killing horizon; this will be addressed in the next
section. The construction mimics the corresponding extension
procedure for the Kerr metric, see
also~\cite[Section~3]{HennigAnsorgCederbaum}
or~\cite{BardeenLesHouches}.

\bigskip

Let $a\in \R$ and let $m>0$ be given by
$$
 m^2 =
 \left(\frac{a_j-a_i}2\right)^2 + a^2
 \;,
$$
set $r_\pm = m \pm \sqrt{m^2-a^2}$. As a first step of the
construction of an extension on $[a_i,a_j] = [a_5,a_4]$ or
$[a_i,a_j] = [a_3,a_2]$ we introduce the usual coordinates
$\tilr$ and $\ttheta$ for the Kerr metric:
\bel{27IV10.2}
   \tilr = \frac{\ri  +\rj }{2} +m
 \;,
 \qquad \ttheta=\cos
   ^{-1}\left(\frac{\rj -\ri  }{a_i-a_j}\right)
   \;,
\ee
with inverse transformation (see, e.g., \cite[(1.133),
p.~27]{MeinelFigures})
\beal{KerrWeyl.1xx}
  \rho & =&
   \sqrt{ {\tilr ^2-2
     m\tilr+a^2} }\, \sin (\ttheta )\equiv
 \sqrt{ {(\tilr-r_-)(\tilr-r_+)}}\, \sin (\ttheta )
   \;,
\\
 \label{KerrWeyl.2xx}
   z & = &\frac {a_i+a_j}2 + { (\tilr -m)} \, \cos ( \ttheta
    )
   \;.
\eea
Note that in the above conventions we have $a_j>a_i$.

In the $(\tilr,\ttheta)$ coordinates the flat metric
$\gamma:=d\rho^2+dz^2$ remains diagonal,
\bean
 \gamma & = &
  \Big(    (\tilr-m)^2- \left(m^2-a^2\right) \cos^2 ( \ttheta )\Big)
\\
 &   & \phantom{xxxxx}\times
\left(\frac{d\tilr^2}{(\tilr -m)^2 -(m^2-a^2)
}+d\ttheta^2\right)
\\
 &= &
{\left(\frac{\rho^2}{\sin^2\ttheta}+(m^2-a^2)\sin^2\ttheta\right)
 }
\left(\frac{\sin^2\ttheta}{\rho^2}d\tilr^2+d\ttheta^2\right)
\\
 &= &
  {\ri  \rj }\left(\frac{d\tilr^2}{{(\tilr-r_-)(\tilr-r_+)}}+d\ttheta^2\right)
\\
 &= &
  {\ri  \rj }\left(\frac{2 \left(\ri \rj +(z-a_i  )
   (a_j -z)-\rho ^2\right)}{\rho ^2 (a_i -a_j )^2}d\tilr^2+d\ttheta^2\right)
 \;,
\eeal{31III10.3}
where the various forms of the metric $\gamma$ have been listed
for future reference.

The essential parameter above is $m^2-a^2$, in the sense that a
change of $m$ and $a$ that keeps $m^2-a^2$ fixed can be
compensated by a translation in $\tilr$, without changing the
explicit form of $\gamma$. The replacement of $\sqrt{m^2-a^2}$
by $-\sqrt{m^2-a^2}$ can be compensated by a change of the sign
of $(\tilr-m)$, which again does not change the explicit form
of $\gamma$.

%
%
%
%
%
We have, near $\rho=0$,  for $a_i<z<a_j$, with error terms not
necessarily uniform over compact sets of $z$,
%
\beal{27IV10.4}
 \gamma_{\tilr\tilr}
  & = &
 \frac{4 (a_i -z)^2 (a_j -z)^2}{\rho^{2}(a_i -a_j )^2}+O(1)
\;,
\\
 \gamma_{\ttheta\ttheta}%
 & = & |(z-a_i)(z-a_j)| + O(\rho^2)
 \;.
\eeal{27IV10.5}

Now, the Black Saturn metric depends upon $\rho$ through
$\rho^2$  only, with the latter being an analytic function of
$\tilr$ and $\ttheta$.   In the new coordinate system all the
metric functions extend analytically across $\{\rho=0, z\in
(a_i,a_j)\}$  \emph{except} $g_{\tilr\tilr}$, which has a first
order pole in $\tilr $ at $\tilr = r_\pm$. In the original
coordinate system we start with $\tilr >r_+$ and it is not
clear whether or not $r=r_-$ can be reached in the analytic
extension, but
 \ptclater{clarify further extensions}
we need to get rid of the pole  at $\tilr = r_+$ in any case.
For this, it is convenient to
\newcommand{\h}{\hat}%
\newcommand{\til}{\tilde }%
%
continue with a general discussion. We consider a coordinate
system $(x^\mu,y)\equiv (x^0,x^i)\equiv (x^0,x^A,y)$, where
$\mu$ runs from $0$ to $n-1$, and we suppose that:
\begin{enumerate}
 \item The metric functions $g_{\mu\nu}$ are defined and
     real analytic near $y=y_0$, except for $g_{yy}$ which
     is meromorphic with a pole of order one at $y_0$.
     \item The determinant of the metric is bounded away
         from zero near $y=y_0$.
 \item There exists a Killing vector field $\xi$ of the
     form
     $$
     \xi = \partial_0 + \alpha^i \partial_i
     \;,
     $$
     for some set of \emph{constants} $\alpha^i$, such that
     all the functions
     $$
     g_{\mu\nu}\xi^\mu
     $$
     vanish at $y=y_0$.
\end{enumerate}

In our case the first condition has just been verified with
$$ y=\tilr\;,\quad y_0=r_\pm \;.
$$

The determinant condition holds by inspection of the metric,
see Tables~\ref{T18VII.1} and \ref{T18VII.0}.

The third condition is verified by a {\sc Mathematica}
calculation, leading to a Killing vector
%
$
 \partial_t + \Omega_{S^3} \partial_\psi
$, where
$$\Omega_{S^3}=-\left(\frac{2 (a_1 - a_2) (a_1 - a_3) (a_2 -
     a_3)}{(a_2 - a_3) (a_1 - a_5) c_1 + (a_3 -a_1) (a_2 - a_5) c_2} + q\right)^{-1}\;,
$$
satisfying the condition on $(a_3,a_2)$, and the Killing vector
$
 \partial_t + \Omega_{S^1\times S^2} \partial_\psi\;,
$ with
$$ \Omega_{S^1\times S^2}=-\left(\frac{2 (a_1 - a_2) (a_1 - a_4) (a_2 -
     a_4)}{(a_2 - a_4) (a_1 - a_5) c_1 + ( a_4 -a_1) (a_2 - a_5) c_2} + q\right)^{-1}\;,
$$
satisfying the condition on $(a_5,a_4)$. A rather lengthy {\sc
Mathematica} calculation shows that the $\Omega$'s are finite
for distinct $a_i$'s.

We introduce new coordinates $(\hat x^\mu, \hat y)\equiv (\hat
x^0, \hat x^A, \hat y) \equiv (\hat x^0, \hat x^i)$ by the
formula
\bel{1XI.1}
 \hat x^0 = x^0  \;,\quad \hat x^i = x^i - \alpha^i x^0
 \;.
\ee
This coordinate transformation has Jacobian one. Writing $g_{\h
\mu \h \nu}$ for $g(\partial_{\h x ^\mu},
\partial_{\h x ^\mu})$, our hypotheses imply that we can write
\bel{20XI.1}
 g_{\h 0 \h \mu} =(y-y_0) \chi_{\h \mu}  \;, \qquad g_{\h y \h y}= \frac{ h }{(y-y_0)}
 \;,
\ee
for some functions $\chi_{\h \mu}$, $h$, all analytic near
$y_0$.

Since the metric functions are now independent of $\h x^0$, the
next coordinate transformation
$$
 d\til x^ 0= d\h x^0 + f(\h y) d\h y \;, \quad \til x^A = \h x^A\;,\quad \til y = \h y
 \;,
$$
again with Jacobian one, does not affect the analyticity
properties of the functions involved. We have
\bean
 g_{\h 0 \h 0} (d\h x^0)^2 + g_{\h y \h y} d\h y^2
  & = &
 (y-y_0) \chi_{\h 0 } \big(d\til x^ 0 - f   d\h y\big)^2 + \frac{h}{  (y-y_0)} d\h y^2
\\
 & = &
 (y-y_0) \chi_{ \h 0} (d\til x^ 0 )^2 - 2(y-y_0)   \chi_{ \h 0} f  d\til x^ 0 d\h y
 \nonumber
\\
 && +
 \frac{h+  (y-y_0)^2 \chi_{\h 0 } f^2 }{  (y-y_0)} d\h y^2
 \;.
\eeal{27IV10.1}
Assume that
%
$$
 \kappa:= -\lim_{y\to y_0}
 \frac {h}{\chi_{\h 0}}
$$
is a positive constant. Keeping in mind that $\chi_{\h 0}$ is
negative while $h$ is positive, and choosing $f$ as
\bel{19VI0.1}
 f= \frac {\sqrt{\kappa}} {y-y_0}
 \;,
\ee
one obtains a smooth analytic extension of the metric through
$y=y_0$, since then  the singularity in \eq{27IV10.1} is
removable; similarly
\beaa
 g_{\h 0 \h i } d\h x^0 d\h x^i
  & = &
 (y-y_0) \chi_{\h i} \big(d\til x^ 0 - f d\h y\big) d\h
 x^i
\\
 & = &
  (y-y_0) \chi_{\h i} d\til x^ 0  d\h
 x^i -     \chi_{\h i} \sqrt{\kappa} d\h y  d\h
 x^i
 \;.
\eeaa
The determinant of the metric in the  coordinate system $\til
x^\mu$ equals that in the original coordinates, and so the
extended metric is Lorentzian near $y=y_0$.

\bigskip

It remains to show that this procedure applies to the BS
metric,
%
%
%
%
%
%
%
%
with
$$
x^0 =t\;, \
 y-y_0:= \tilr - r_+\;, \ (x^A)=(\varphi,\psi,\ttheta)
\ \;,
$$
where $\tilr$ and $\ttheta$ have been defined in \eq{27IV10.2}.
We have
$$
\tilr-r_+ = \frac{(a_j-a_i)\rho^2}{4 (a_j-z)(z-a_i)} + O(\rho^4)
 \;,
$$
hence
$$
(\tilr-r_- ) \sin^2 \ttheta= \frac{4 (a_j-z)(z-a_i)}{(a_j-a_i) } + O(\rho^2)
 \;,
$$
with the error term \emph{not} uniform in $z$ near the end
points.  On $(a_5,a_4)$ or  on $(a_3,a_2)$ one needs to
calculate the limits
\bean
 h|_{\tilr =r_+}
  &=&
   \lim_{\rho\to 0} \frac {H_x k^2 P}{(\tilr -r_-) \sin^2 \ttheta} \times \lim_{\rho\to 0} (\rho^2 \gamma _{\tilr
   \tilr})
 \;.
\eeal{31III10.5}
Letting $\Omega=\Omega_{S^1\times S^2}$ on $(a_5,a_4)$,
respectively $\Omega=\Omega_{S^3}$ on $(a_3,a_2)$, one further
needs
\bean
 \chi_{\h 0}|_{\tilr =r_+}
  &=&
    \lim_{\rho\to 0} \big(\rho^{-2}  g(\partial_t + \Omega \partial_\psi,\partial_t + \Omega \partial_\psi)(\tilr -r_-) \sin^2
    \ttheta \big)
 \;.
\eeal{31III10.6}
A surprisingly involved {\sc Mathematica} calculation shows
that at $\rho=0$ the quotient ${h}/{\chi_0}$ equals, up to
sign,
$$
 \frac{(a_4-a_5) (2 (a_1-a_2) (a_1-a_4) (a_2-a_4)+((a_2-a_4) (a_1-a_5) c_1+(a_4-a_1) (a_2-a_5) c_2) q)^2}{8 (a_1-a_2)^2 (a_2-a_4) (a_3-a_4)^2 (a_4-a_1)}
$$
on $(a_5,a_4)$, and
$$
\frac{(a_3 - a_5) (2 (a_1 - a_2) (a_1 - a_3) (a_2 -
         a_3) + ((a_2 - a_3) (a_1 - a_5) c_1 + (a_3-a_1) (a_2 -
            a_5) c_2) q)^2}{8 (a_1 - a_2)^2 (a_2 - a_3) (a_3-a_1) (a_3 -
      a_4)^2}
$$
on $(a_3,a_2)$.  As those limits are constants, we have
verified that, within the current range of parameters, the
Black Saturn metric can be extended across two non-degenerate
Killing horizons.

\subsection{Intersections of axes of rotations and horizons}
\label{ssIntAxHor13VII}

It follows from \eq{27IV10.2} that
\bea
 \ri  &= & 
\tilr-r_+ +\frac{a_j-a_i}{2}( \cos\ttheta+1)
 \;,
  \label{28IV10.1}
\\
 \rj  &= & \tilr-r_+ +\frac{a_j-a_i}{2}(1- \cos\ttheta)
 \;,
\\
 \mu_i
&= & (\tilr-r_+) (1-\cos\ttheta)
 \;,
\\
 \mu_j
 &= & (\tilr-r_-) (1-\cos\ttheta)
 \;,
\eeal{28IV10.2}
%
so that  $\mu_i$, $\mu_j$, $R_i$ and $R_j$  are smooth
functions of $\tilr$ and
$\cos\ttheta$.%
\footnote{It should be kept in mind that $\cos \ttheta$ is a
smooth function on the sphere, but $\sin \ttheta$ is not.}
Furthermore, it follows from \eq{KerrWeyl.1xx} that the
function $\rho^2$ is a smooth function of $\tilr$ and of
$\sin^2\ttheta=1-\cos^2 \ttheta$, similarly $z$ is smooth in
$\cos \ttheta$ by \eq{KerrWeyl.2xx}, which implies that  the
remaining $\mu_\ell$'s (compare~\eq{19VI0.2}-\eq{19VI0.5}) are
smooth in $\tilr$ and $\cos\ttheta$.

Now, consider any rational function, say $W$, of the $\mu_i$'s
and $\rho^2$, which is bounded near $\tilr=r_+$, $\ttheta=0$.
Boundedness implies that any overall factors of $\tilr - r_+$
in the denominator of $W$ are cancelled out by a corresponding
overall factor in the numerator, leaving behind a denominator
$d(\tilr,\ttheta)$ which can be written in the form
\bean
d(\tilr,\ttheta) =\mathring f(\cos\ttheta) + (\tilr-r_+)\mathring g(\tilr, \cos\ttheta)
 \;,
\eeal{2VII10}
for some functions $\mathring f $ and $\mathring g$ which are
smooth in their respective arguments. \emph{If}
$$
 d(\tilr=r_+,0)\equiv \mathring f(1)
$$
does \emph{not} vanish at $\ttheta=0$, \emph{then} the
denominator $d$ is bounded away from zero near  $\tilr=r_+$ and
$\ttheta=0$.
This in turn implies that $1/d$ is smooth in a
neighbourhood of the point concerned, and therefore so is $W$.

An identical argument applies at $\ttheta=\pi$.

This reasoning does not seem to apply to $\omega_\psi$, because
of the square roots there. However, as mentioned in
Appendix~\ref{ssMetrCoeff13VI}, these appear in the form
$$
 \sqrt{\frac{M_0M_1}{G_x}}\;,
 \quad
 \sqrt{\frac{M_0M_2}{G_x}}\;,
 \quad
 \sqrt{\frac{M_1M_4}{G_x}}\;,
 \quad
 \sqrt{\frac{M_2M_4}{G_x}}
 \;.
$$
One checks that the expressions under the square root are
squares of  rational functions of the $\mu_i$'s, and of
$\rho^2$, and so the metric functions involving $\omega_\psi$
are also rational functions of the $\mu_i$'s and $\rho^2$.

Since we have already shown that the suitably reduced
denominators
of all the scalar products $g(X,Y)$, where $X,Y\in
\{\partial_t,\partial_\psi,\partial_\varphi\}$, have no zeros
at the axis points $\rho=0$, $z=a_i$, we conclude that the
corresponding metric coefficients are analytically extendible,
by  allowing $\tilr$ to become smaller than $r_+$, including
near the intersections of axes of rotation with the Killing
horizons.

One similarly establishes analytic extendibility of $ g_{\tilde
t \tilde y}$:
%
%
\beaa
 g_{\tilde y\tilde t}&=&-\frac{(g_{tt}+2g_{t\psi}\Omega+g_{\psi\psi}\Omega^2)\sqrt \kappa}{\tilr - r_+}
 \;.
\eeaa
Here we have already verified that $
g_{tt}+2g_{t\psi}\Omega+g_{\psi\psi}\Omega^2 $ is an analytic
function of $\tilr$ and $\cos\ttheta$, and  extendibility of
$g_{\tilde y\tilde t}$ readily follows from the fact that
$\Omega$ has been chosen so that this function vanishes at
$\tilr=r_+$.

Finally,  $ g_{\tilde y \tilde y}$ is given by the formula
%
%
\beal{13VII0.1}
 g_{\tilde y\tilde y}&=& \frac{ \sqrt \kappa g_{\tilde y\tilde t}  +(\tilr-r_+) g_{\tilde
 r\tilde r}}{\tilr-r_+}\;.
\eea
To analyse this metric function,
by a {\sc Mathematica} calculation we verified that the reduced
denominator of $(\tilr-r_+)g_{\tilde r\tilde r}$ does not
vanish at $\tilr=r_+$, and hence this function extends across
$\tilde r = r_+$ as an analytic function of $\tilr$ and
$\cos\ttheta$. Keeping in mind that the same has already been
established for $\sqrt \kappa g_{\tilde t\tilde y}$, we find
that the numerator of \eqref{13VII0.1} extends across $\tilde r
= r_+$  as an analytic function of $\tilr$ and  $\cos\ttheta$.
Analytic extendibility of  $ g_{\tilde y\tilde y}$ follows
again from standard factorisation properties of such functions.

We next analyse $g_{\ttheta\ttheta}$  near  $\rho=0$, $z=a_4$.
Now,
\beaa
 g_{\ttheta\ttheta} &=& H_x k^2 P \gamma _{\ttheta\ttheta} = g_{\rho\rho}   \ri  \rj
 \;,
\eeaa
and we need to understand the behaviour of the functions above
near $\tilr =r_+$, $\ttheta\in \{0,\pi\}$. 
For $\ell \ne 5$ we have
\beal{19VI0.2}
 \mu_\ell \mu_5 + \rho^2 &=& \left((\tilr - r_-)\sin^2\theta +\mu_\ell (1-\cos\ttheta)\right)
  (\tilr - r_+)
 \;,
\eea
and since
\bel{19VI0.5}
 \mu_1 =\frac{ \rho^2}{R_1+z-a_1}\approx \frac{ \rho^2}{2(a_4-a_1)}
\ee
near $\rho=0$, $z=a_4$,  for $\ell =2,3$ we can write
\bea
 \mu_\ell \mu_4 + \rho^2 &=& \left(\tilr - r_+ +\frac{\mu_\ell}{1+\cos\ttheta}\right)
  (\tilr - r_-)\sin^2 \ttheta
 \;,
\\
 \mu_\ell \mu_5 + \rho^2 &=& \left(\tilr - r_- +\frac{\mu_\ell}{1+\cos\ttheta}\right)
  (\tilr - r_+)\sin^2 \ttheta
 \;,
\\
 \mu_4 \mu_5 + \rho^2 &= &  \frac{2 (\tilr - r_-)}{1+\cos\ttheta}
  (\tilr - r_+)\sin^2 \ttheta
 \;,
\\
 \mu_1 \mu_\ell + \rho^2 &\approx & \left(\frac{a_\ell-a_4}{a_4-a_1}+1\right)\rho^2
  \nonumber
\\
   &= &
   \frac{a_\ell-a_1}{a_4-a_1} (\tilr-r_-)(\tilr-r_+)\sin^2
  \ttheta
 \;.
\eea
Finally, for $\ell=1,4,5$,
\bea
 \mu_1 \mu_\ell + \rho^2 &\approx & \rho^2 =(\tilr-r_-)(\tilr-r_+)\sin^2
  \ttheta
 \;.
\eeal{19VI0.3}
Encoding this behaviour into a {\sc Mathematica} calculation,
one finds that $g_{\ttheta\ttheta}$ is uniformly bounded in a
neighbourhood of $r=r_+$, $\cos \ttheta\in \{\pm 1\}$, with
non-vanishing value of the denominator as needed above. This
establishes smoothness. Similarly
$g_{\varphi\varphi}/\sin^2\ttheta$ is smooth near those points.

Now, away from, and near to, the event horizons, the map
$(\rho,z) \mapsto (\tilr,\ttheta)$ is a smooth coordinate
transformation. From what has been already established,  the
two-dimensional metric
\bel{17VI0.1}
 g_{\ttheta\ttheta} d\ttheta^2 + g_{\varphi\varphi} d\varphi^2
\ee
is thus a smooth metric for $\tilr>r_+$, $\tilr$ close to
$r_+$, in particular there is no conical singularity at the
rotation axis for $\partial_\varphi$ in this region. But the
arguments just given show that this metric extends smoothly
across $\tilr=r_+$, which finishes the proof of smoothness of
the whole metric up-to-and-beyond the horizon near $\tilr=r_+$,
$\ttheta=0$.

A similar analysis applies near $a_5$, $a_3$ and $a_2$; in this
last case, one considers the two-dimensional metric
$$
 g_{\ttheta\ttheta} d\ttheta^2 + g_{\psi\psi} d\psi^2
$$
instead of \eq{17VI0.1}.

\subsection{Event horizons}
 \label{ss24IV10.1}

Consider the manifold, say $\mcM$, obtained by adding to the
region $\tilr>r_+$ those points in the region $r_-<\tilr$ for
which the metric is smooth and Lorentzian. Then the region
$r_-<\tilr\le r_+$ is contained in a black hole region in the
extended space-time, which can be seen as follows: Note, first,
that $g^{yy}$
vanishes at $\mcH:=\{\tilr = r_+\}=\{y = y_0\}$, which shows
that $\mcH$ is the union of two null hypersurfaces. On each
connected component of $\mcH$ the corresponding Killing vector
$X=\partial_t + \Omega
\partial_\psi$ is timelike future pointing for $y>y_0$ close to
$y_0$, and so by continuity $X$ is future pointing on $\mcH$.
This implies that $\mcH$ is \emph{locally achronal} in the
extended space-time: if a future directed timelike curves
crosses $\mcH$ through a point $p\in \mcH$, it does so towards
that side of $T_p\mcH$ which contains the component of the set
of causal vectors at $p$  containing $X$. Since $\mcH$ is a
(closed) separating hypersurface in $\mcM$, this implies that
any timelike curve can cross $\mcH$ only once. From what has
been said it follows that the region $r_-<\tilr\le r_+$ is
contained in a black hole region of $(\mcM,g)$.

In particular we have shown that the black hole region is not
empty. A standard argument (compare~\cite[Section~4.1]{CC})
shows that $\mcH$ coincides with the black hole event horizon
in $\mcM$. Note that this is true independently of stable
causality of $(\mcM,g)$, or of stable causality of the d.o.c.
in $(\mcM,g)$.

Some more work is required to add the bifurcation surface of
the horizon, a general procedure how to do this is described
in~\cite{RaczWald2}.
 \ptclater{expand upon this?}

\subsection{The analysis for $c_2 = 0$}
 \label{Schoicec2zero}

We turn our attention now to  the Black Saturn solutions with
$c_2 = 0$, where the formulae simplify sufficiently to allow a
proof of stable causality of the d.o.c.

\begin{figure}
\begin{center}
 \includegraphics[scale=.4]{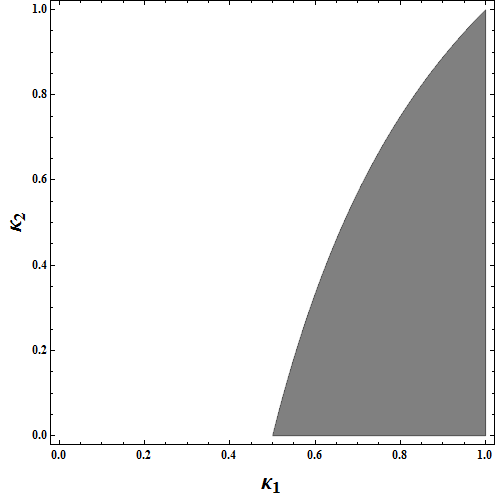}
\end{center}
 \caption{\label{Fkappas}The variable $\kappa_1$ runs along the horizontal axis, while $\kappa_2$ runs along the vertical one.
The inequality \eq{30III10.4} corresponds to the shaded region.}
\end{figure}

First note that (\ref{c_2conical}) implies that the condition
$c_2 = 0$ leads to $c_1 \neq 0$ as the only restriction on
$c_1$. However, it implies a fine-tuning of the parameters
$a_i$. One may easily check that the minus sign solution for
$c_2$ cannot vanish if the ordering  \eq{21IV.41.1} of the
$a_i$'s is assumed. However the plus sign solution may lead to
the vanishing $c_2$ under certain additional conditions. Namely
the resulting equation
\beaa   \sqrt{({a_3  }-{a_1 }) ({a_2  }-{a_4  }) ({a_2  }-{a_5
}) ({a_3 }-{a_5 })}
 =  ({a_2  }-{a_1 }) ({a_3  }-{a_4  })
 \;,
\eeaa
quadratic in $a_5$, may always be solved for $a_5 = a_5
(a_1,a_2,a_3,a_4)\in \R$; the condition that $0<a_5<a_4$ is
then equivalent to
\bel{30III10.3}
 a_4   < (a_2  ^2 + a_1  a_3 -
 2 a_2   a_3  )/(a_1 - a_3  )
 \;.
\ee
This is more transparent in terms of the variables $\kappa_i\in
[0,1]$ defined by \eq{21IV.43}, as then \eq{30III10.3} becomes
\bel{30III10.4}
 \kappa_1> \frac 1 {2-\kappa_2}
 \;,
\ee
see Figure~\ref{Fkappas}.
In the further analysis one should keep in mind that $a_5$ is
no more an independent parameter.

Notice that $c_2 = 0$ implies $q=0$ and $k=1$.

\subsubsection{Smoothness at the axis}
 \label{ssAor25III10}

Smoothness of the Black Saturn solution for $\rho > 0$, proved
in Section \ref{SRegularity13VI}, holds also for the $c_2 = 0$
case, hence only the analysis on the axis of rotation needs
separate attention. We shall proceed in the same way as in
Section \ref{ssAor13VII}.

We start  with an analysis of the behaviour of $g_{\psi\psi}$
on the axis. For $z<a_1 $ it may be written as a rational
function
%
\beaa
 -\frac{2 ({a_1 }-{a_3  })^2 ({a_2  }-z) (z-{a_2  }) (z-{a_4  })
 (z-{a_5 })+{c_1}^2 ({a_1 }-{a_2  })^2 ({a_1 }-{a_5 })^2 ({a_3  }-z)}{({a_1}-{a_3  })^2 ({a_1 }-z) (z-{a_2  }) (z-{a_4  })}
 \;.
\eeaa
To avoid the singularity at $z = a_1 $ we need to fix $c_1$ as
to have a finite limit. Miraculously, this condition leads to
the same formula $c_1$ as obtained in section 2 for $c_2 \neq
0$. This is somewhat unexpected, since  we have set $c_2$ to
zero as an alternative to fixing $c_1$. With this choice of
$c_1$ regularity on the axis of many metric functions has
already been established, and we would be done if not for the
fact  that some of the formulae derived so far involve explicit
inverse powers of $c_2$. So it is necessary to repeat the
analysis at the axis from scratch.

Several formulae are much simpler now. For instance, one checks
that in the region $a_1 < z \leq a_5$ on the axis
$g_{\psi\psi}$ is given by the same formula as for $z <
a_1$.
Hence we conclude, that $g_{\psi\psi}$ is smooth
and bounded for $\{ \rho = 0 , \, z < a_5\}$.

In the subsequent axis interval, $a_5 < z < a_4$,
$g_{\psi\psi}$ is a rational function with denominator
%
\beaa 2 ({a_1 }-{a_2  })^2 ({a_1 }-{a_4  })^2 ({a_3  }-z) ({a_5
}-z)-{c_1}^2 ({a_1 }-{a_5 })^2 ({a_2  }-z)^2 ({a_4  }-z)\;, \eeaa
which cannot vanish, being a sum of two negative terms. At both
end points of the investigated interval one of the terms in
non-zero, which shows boundedness.

Moving further to the right we obtain a simple formula for
$g_{\psi\psi}$:
%
\bel{27III10.1}
  \frac{2(a_1  - z) (a_2   - z)}{(a_5  - z)}
 \;,
\ee
which immediately implies continuity for $a_4 \leq z \leq a_3$.
We note that this is strictly positive, and therefore near that
axis interval $g_{\psi\psi}$ is strictly positive as well.

In the region $a_3 < z < a_2$ the denominator of $g_{\psi\psi}$
is more complicated:
%
\beaa (a_1  - a_5 )^2 c_1^2 (a_2   - z)^2 (a_3   - z) - 2 (a_1 -
a_2  )^2 (a_1  - a_3  )^2 (a_4   - z) (a_5  - z), \eeaa 
but  does not vanish, being a strictly negative  sum of two
non-positive terms.

In the region $z > a_2$ for vanishing $c_2$ the function
$g_{\psi\psi}$ is proportional to $q^2$. Since $c_2 = 0$
implies $q=0$, we conclude that $g_{\psi\psi}$ vanishes for $z
> a_2$, as already seen for general values of $c_2$ in any
case.

The analysis of $g_{tt}$ is similar. For $\rho = 0$ and $z <
a_5$ the metric function $g_{tt}$ is a simple rational
function,
\bel{24III10.2}
 \frac{(a_1  - z) (z -a_3)}{(z -a_2 ) (z -a_4 )}
 \;,
\ee
which is clearly continuous in the region $z \leq a_5$. For
$a_5 < z < a_4  $ the denominator of $g_{tt}$ reads
%
\bel{24III10.4}
 (a_1 - a_5 )^2 c_1^2 (a_2   - z)^2 (a_4   - z) + 2 (a_1
 - a_2  )^2 (a_1  - a_4  )^2 (a_3   - z) (z -a_5)\;.
\ee
with both terms manifestly positive in the region $a_5  \le z
\le a_4  $. We conclude that $g_{tt}$ is smooth on $a_5  < z <
a_4 $, bounded on  $a_5  \leq z \leq a_4  $.

Next, for $a_4   < z < a_3  $ the denominator of $g_{tt}$
reads
%
\beaa 2 (a_1  - a_2  )^2 (a_1  - z) ( z -a_2) (z -a_5 )\;,
\eeaa
thus it cannot vanish for $a_4   \leq z \leq a_3  $. Moving
further to the right we find the denominator of $g_{tt}$
%
\beaa (a_1  - a_5 )^2 c_1^2 (a_2   - z)^2 (a_3   - z) + 2 (a_1 -
a_2  )^2 (a_1  - a_3  )^2 (a_4   - z) ( z-a_5)
 \eeaa
as a sum of manifestly negative terms on $a_3   < z < a_2  $.
Also the end points are singularity-free. Finally, for $z
> a_2$ $g_{tt}$ equals
%
\bel{24III10.3} \frac{(z -a_2 ) (z-a_4)}{(a_1  - z)
(z-a_3)}
 \;,
\ee
hence it is continuous. This proves directly absence of
singularities for $g_{tt}$ on the axis in the case of vanishing
$c_2$.

The analysis of $g_{t\psi}$ can be carried out along the same
lines, and is omitted.

\subsubsection{Causality away from the axis}
 \label{ssCausal25III.1}

We have not been able to establish non-existence of closed
timelike curves for a general Black Saturn solution, though we
failed to find any in a numerical search, see
Appendix~\ref{nr}. However, if one imposes the condition $c_2 =
0$ the metric formulas simplify sufficiently to allow a direct
analysis. Indeed, the explicit formula for $g_{\psi\psi}$ in
the case of vanishing $c_2$ (and consequently $q = 0$) reads
%
\beaa \frac{{\mu_1} {\mu_2} {\mu_5} \left(\rho ^2 \left({c_1}^2
M_1 +M_0 \right)^2-4 {c_1}^2 M_0  M_1 {R_1}^2\right)} {\rho ^2
\left({c_1}^2 M_1 +M_0 \right) \left(M_0  {\mu_1}^2-{c_1}^2 M_1
\rho ^2\right)} =: \frac{f(c_1^2)}{g(c_1^2)}\;. \eeaa
Outside the axis ($\rho >0$) the ordering of $\mu_i$'s is the
same as those of $a_i$'s and all the functions $M_i$ are
strictly positive. Both the numerator and denominator of
$g_{\psi\psi}$ can be regarded as quadratic functions of
$c_1^2$. Let us first investigate the possible zeros of the
denominator:
%
\beaa
 g(c_1^2) = 0 \Rightarrow c_1^2 = -\frac{M_0}{M_1} \text{
 or } c_1^2 =\frac{M_0 \mu_1^2}{M_1 \rho^2}\;.
\eeaa
Clearly only the second one is relevant since the first one
would lead to an imaginary coefficient $c_1$. On the other
hand, the  equation $f(c_1^2) = 0$ has two solutions:
\beaa \cpm  = \frac{{M_0} \left(2 {R_1} \left({R_1}\pm
\sqrt{{R_1}^2-\rho^2}\right)-\rho ^2\right)}{{M_1} \rho ^2}\;.
\eeaa
To make this result more transparent let us express $R_1$ in terms of $\mu_1$ and $\rho$
\beaa
R_1 = \frac{\mu_1^2+\rho^2}{2 \mu_1}
 \;.
\eeaa
Then $\cpm $ may be written as
\beaa \cpm  = \frac{M_0}{{M_1} \rho ^2} \,
\frac{(\mu_1^2+\rho^2) \left(\mu_1^2+\rho^2 \pm |\mu_1^2 -
\rho^2| \right) - 2 \mu_1^2 \rho^2}{2 \mu_1^2}\;. \eeaa
From the explicit form of $\mu_1$ one can easily see that
$\text{sign}(\mu_1^2 - \rho^2) = \text{sign}(a_1-z)$, thus we
have:
\beaa \text{for } z \le a_1  & \displaystyle \csm  = \frac{M_0
\mu_1^2}{{M_1} \rho ^2}\;,
& \csp  = \frac{M_0 \rho ^2}{{M_1} \mu_1^2}\;, \\
\text{for } z \ge a_1  &  \displaystyle \csm  = \frac{M_0 \rho
^2}{{M_1} \mu_1^2}\;, & \csp  = \frac{M_0 \mu_1^2}{{M_1} \rho
^2}\;. \eeaa
We see that in both regions one of the zeros $\cpm $ of the
numerator cancels the zero of the denominator, which provides
an alternative explicit proof of regularity of $g_{\psi\psi}$
for $\rho>0$. Moreover we find:
\beaa
   g_{\psi\psi} &= &   \frac{{\mu_1} {\mu_2} {\mu_5} M_1} {\rho ^2
 \left({c_1}^2 M_1 +M_0 \right)  } \left( \frac{M_0 \rho ^2}{{M_1}
    \mu_1^2} -c_1^2 \right)
 = \frac{{\mu_2} {\mu_5}  \left(  {M_0 \rho ^2}  -c_1^2 {M_1}
    \mu_1^2\right)} {\rho ^2{\mu_1}
 \left({c_1}^2 M_1 +M_0 \right)  }
 \;.
 \eeaa
Keeping in mind that the parameter $c_1$ has been fixed to
guarantee the regularity on the axis, to obtain a sign for
$g_{\psi\psi}$ for $\rho>0$ it remains to show that the
equality
\beaa
 c_1^2 = \frac{M_0 \rho ^2}{{M_1} \mu_1^2}
\eeaa
can never be satisfied away from the axis. For this, we shall
make use of the formula (\ref{c_1.2s}) expressing  $c_1^2$ in
terms of $\mu_i$'s. By subtracting the two formulae for $c_1^2$
we obtain
\beaa -\frac{{\mu_5} ({\mu_3}-{\mu_1}) \left({\mu_1}
{\mu_4}+\rho ^2\right)}{{\mu_1}^4 {\mu_3} {\mu_4}
({\mu_1}-{\mu_2})^2 ({\mu_1}-{\mu_5})^2 \left({\mu_1} {\mu_5}+\rho ^2\right)} \; \times
 \phantom{xxxxxxxxxxxxxx}
\\
 \Big( {\mu_1}^3 ({\mu_1}-{\mu_2})^2 ({\mu_1}-{\mu_4}) ({\mu_1}-{\mu_5}) \left({\mu_1} {\mu_3}+\rho
^2\right)
 \phantom{xxxxxxxxxxx}
\\
 +({\mu_1}-{\mu_3}) \left({\mu_1} {\mu_2}+\rho ^2\right)^2
 \left({\mu_1} {\mu_4}+\rho ^2\right) \left({\mu_1} {\mu_5}+\rho ^2\right)
\Big) =0
 \;.
\eeaa
The overall multiplicative coefficient in the first line is
strictly negative, whereas the term in parenthesis across the
second and third lines is a polynomial in $\rho$ with
coefficients that can be written in the following, manifestly
negative form
\beaa & &
 \mu_1^5 \left( (\mu_1-\mu_2)^2 \mu_3 (\mu_1-\mu_4) +
 \mu_5 \left( \mu_2^2 (\mu_4-\mu_3) + (\mu_1-\mu_4) \mu_3 (\mu_2
 + (\mu_2 - \mu_1) ) \right) \right)
\\
& &
 +
 \rho^2 \mu_1^3 \Big( \mu_2 (\mu_1 - \mu_3) (2 \mu_4 \mu_5 +
 \mu_2 (\mu_5+\mu_1))
\\
& &
 \phantom{ \rho^2 \mu_1^3 \Big(}
 + (\mu_4-\mu_1) \left(
 \mu_2^2(\mu_5 - \mu_3) + \mu_1 ( (\mu_1 - \mu_2) - \mu_2 )(\mu_5
 - \mu_1) \right) \Big)
\\
& &
 +
 \rho^4 \mu_1^2 (\mu_1-\mu_3) \left( \mu_2^2+\mu_4 \mu_5 + 2
 \mu_2 (\mu_4+\mu_5) \right)
\\
& &
 +
 \rho^6 \mu_1 (\mu_1-\mu_3) (2 \mu_2 + \mu_4 + \mu_5)
\\
& &
 +
 \rho^8 (\mu_1 - \mu_3)
 \;.
\eeaa
It follows that $g_{\psi\psi}>0$ for $\rho>0$ when $c_2=0$.

It turns out that an alternative simpler argument for
positivity can be given as follows: Using \eq{c_1.2s} we may
write $g_{\psi\psi}$ in terms of $\mu_i$ and $\rho$. The
functions $\mu_i$ satisfy the same ordering as $a_i$
\eq{21IV.41.1} (see \eq{21IV.45}). The strict version of the
ordering \eq{21IV.41.1} implies a strict ordering of the
$\mu_i$'s for $\rho>0$. Assuming that, we may make the
positivity of $g_{\psi\psi}$ explicit by expressing it in terms
of the positive functions
\bel{29IV10.21}
 \mbox{$\Delta_{51}=\mu_5-\mu_1$,
 $\Delta_{45}=\mu_4-\mu_5$, $\Delta_{34}=\mu_3-\mu_4$, and
 $\Delta_{23}=\mu_2-\mu_3$.}
\ee
 The numerator and denominator of $g_{\psi\psi}$ are
polynomials in $\Delta_{ij}$, $\mu_1$ and $\rho$, the explicit
form of which is too long to be usefully exhibited here. By
inspection one finds that all coefficient of these polynomials
are positive, and since the $\Delta_{ij}$'s, $\mu_1$ and $\rho$
are positive, both the numerator and denominator of
$g_{\psi\psi}$ are positive.

\subsubsection{Causality on the axis}
 \label{ss16IV10.2}

We turn now our attention to the axis. By continuity, we know
that $g_{\psi\psi}$ at $\rho=0$ is non-negative. It therefore
suffices to exclude zeros of $g_{\psi\psi}|_{\rho=0}$.
Equivalently, whenever we find a manifestly non-zero value of
$g_{\psi\psi}(0,z)$, we know that this value cannot be
negative.

Now, at $\rho=0$ and for $z<a_1$ we replace $z$ by
$w:=z-a_1<0$, and find that $g_{\psi\psi}$ there is a rational
function with denominator
$$
 (a_1  -a_3  ) (a_1  -a_2  +w) (a_1  -a_4  +w)
\;,
$$
which is seen to be strictly negative for $w\le 0$. On the
other hand, the numerator is a third-order polynomial in $w$:
\beaa
 &&
  2 (a_2  -a_1  ) \times \Big(3 a_1  ^3-a_1  ^2
(a_2  +2 (2
   a_3  +a_4  +a_5 ))
\\
 && \phantom{xxx}+a_1   (2 a_2   a_3  +3 a_3
   (a_4  +a_5 )+a_4   a_5 )-a_2   (a_3
   (a_4  +a_5 )-a_4   a_5 )-2 a_3   a_4   a_5 \Big)
\\
 &&
 +2 w (a_3  -a_1  ) \left(6 a_1(a_1-a_2)   -3
a_1   (  a_4  +a_5 )+a_2  ^2+2 a_2   (a_4  +a_5 )+a_4
   a_5 \right)
\\
 &&
+2 w^2
   (a_3  -a_1  ) (4 a_1  -2 a_2  -a_4  -a_5 )
\\
 &&
+2 w^3
   (a_3  -a_1  )
 \;.
\eeaa
Unless explicitly indicated otherwise, the remaining analysis
uses the choice of origin and scale given by $a_1=0$ and
$a_2=1$, which involves no loss of generality for checking the
sign of $g_{\psi\psi}$. The above reduces then to
\beaa
 && -2 ((2 a_3-1) a_4 a_5+a_3
   (a_4+a_5))+2 a_3 w
(a_4 a_5+2
   (a_4+a_5)+1)
\\
 &&
-2 a_3 w^2 (a_4+a_5+2) +2 a_3 w^3
 \;.
\eeaa
Each monomial in the above polynomial is manifestly strictly
negative for $w<0$, except perhaps for the zero-order term.
However, when $c_2=0$, in the current choice of scale we
necessarily have $a_3> 1/2$ by \eq{30III10.3}, which makes
manifest the negativity of the zero-order term as well. Hence
$g_{\psi\psi}|_{\rho=0}>0$ for $z\le a_1$.

The interval $(a_1,a_5)$ requires more work, and will be analysed
at the end of this section.

For $z\in (a_5,a_4)$ we obtain
$$
g_{\psi\psi}|_{\rho=0}=-\frac{2 a_4 (z-1) (a_3-z)}{a_3
\left(a_4 (a_5 (z-2)+1)-a_5
   (z-1)^2\right)+a_4 (a_5-z)}
 \;,
$$
which has no zeros in $[a_5,a_4]$, and thus is positive there.

Positivity on $[a_4,a_3]$ follows already from
\eqref{27III10.1}.

For $z\in (a_3,a_2)$ we obtain
$$
g_{\psi\psi}|_{\rho=0}=-\frac{2 a_3 (z-1) (a_4-z)}{a_3
(a_4 a_5 z-2 a_4
   a_5+a_4+a_5-z)-a_4 a_5 (z-1)^2}
\;,
$$
which again has no zeros in $[a_3,a_2]$, and hence is positive
there.

We already know that $\{\rho=0,z>a_2\}$ is a regular axis of
rotation for $\partial_\psi$, so there are no causality
violations there associated with $\partial_\psi$.

We consider now the interval $(a_1,a_5)=(0,a_5)$. There we find
$$
 g_{\psi\psi}|_{\rho=0}=\frac{f}{a_3 (-1 + z) (-a_4 + z)}
 \;,
$$
with
$$
 f:=\left(a_3 \left(a_4 \left(-(a_5+2) z+2
a_5+z^2+1\right)+(z-1)^2
   (a_5-z)\right)-a_4 a_5\right)
 \;.
$$
Suppose that there exists $z$ in this interval such that $f$
vanishes for some $0<a_5<a_4<a_3<1$. Since $f$ does not change
sign, this can only occur if at this value of $z$ we also have
$$
 \partial_{a_5} f=
 \partial_{a_4} f=
 \partial_{a_3} f= 0
\;.
$$
Now,
$$
 \partial_{a_4} f= 2 (-a_5 + a_3 (-a_5 (-2 + z) + (-1 + z)^2)) \;,
$$
$$
 \partial_{a_5} f = 2 (-a_4 + a_3 (-a_4 (-2 + z) + (-1 + z)^2))
 \;.
$$
The resultant of these two polynomials in $z$ is
$$
16 (a_3-1)^2 a_3^2 (a_4-a_5)^2
 \;,
$$
which is strictly positive in the region of interest, hence
$g_{\psi\psi}$ is also strictly positive on $\{\rho=0,
z\in(a_1,a_5)\}$.

An alternative argument for positivity at $\rho=0$ can be given
as follows: Since all terms in the numerator and denominator
are non-negative one needs to check zeros of the numerator and
denominator. The analysis is done separately on each interval
$(a_i,a_j)$. Before passing to the limit $\rho=0$, for $z> a_i$
the functions $\Delta_{ij}$  (as  defined in \eq{29IV10.21},
and which necessarily vanish at $\rho=0$) are replaced by
positive functions $\hat\Delta_{ij}$ such that
$\Delta_{ij}=\rho^2\hat\Delta_{ij}$. Furthermore we introduce
$\mu_1=\rho^2\hat\mu_1$ for $z>a_1$. Substituting these
expressions in respective intervals of $z$, cancelling common
factors and taking the limit $\rho\rightarrow 0$ one obtains
expressions for the numerator and the denominator of
$g_{\psi\psi}$ at $\rho=0$. These expressions turn out to be
polynomials \emph{with all coefficients
positive}.
For example for $z\in(a_4,a_3)$
we obtain the manifestly positive expressions
$$
 g_{\psi\psi}|_{\rho=0}
 =\frac{(\Delta_{23}+\Delta_{34}) (\hat\Delta_{51}+\hat\mu_1) (1+(\Delta_{23}+\Delta_{34}) \hat\mu_1)^2}{\hat\mu_1 (1+(\Delta_{23}+\Delta_{34}) \hat\mu_1)^2}$$
and for $z\in(a_3,a_2)$
$$
 g_{\psi\psi}|_{\rho=0}
 =\frac{\Delta_{23} (\hat\Delta_{34}+\hat\Delta_{45}+\hat\Delta_{51}) (\hat\Delta_{51}+\hat\mu_1) (1+\Delta_{23} \hat\mu_1)^2}{\hat\mu_1 (\hat\Delta_{45}+\hat\Delta_{51}+\hat\Delta_{34} (1+\Delta_{23} \hat\mu_1)^2+\Delta_{23} (\hat\Delta_{45}+\hat\Delta_{51}) \hat\mu_1 (2+\Delta_{23} (\hat\Delta_{51}+\hat\mu_1)))}
 \;.
$$
It turns out that the
denominator never vanishes and the numerator vanishes, as
expected, only at the axis of rotation of $\partial_\psi$
($z\geq a_2$).

\subsubsection{Stable causality}
 \label{ss16IV10.1}

Using  \eqref{25III10.1x},
$$
 g(\nabla t, \nabla t) = g^{tt} = - \frac{g_{\psi\psi}}{G_y}
 \;,
$$
we conclude from what has been said so far and from
Table~\ref{T18VII.1} that $t$ is a time-function on
\bel{30III10.1}
 \{\rho>0\}\cup\{\rho=0, z\not \in [a_5,a_4]\cup
[a_3,a_2]\}
 \;,
\ee
except perhaps for $\rho=0$, $z>a_2$. There we find
$$
 \lim_{\rho \to 0} \frac{g_{\psi\psi}}{\rho^2}= \frac{(z-a_1)}{2 ( z-a_2) (z-a_5)}
 \;,
$$
which ends the proof of \emph{stable causality} of the region
\eqref{30III10.1} when $c_2=0$. (The blow-up at $z=a_2$ appears
surprising at first sight, but turns out to be compatible with
a smooth axis of rotation, as clarified in
Section~\ref{ssIntAxHor13VII}; compare also \eqref{24III10.1}.)

\bigskip

\appendix

\section{The metric}
 \label{s15VII0.1}

\subsection{The metric coefficients}
\label{ssMetrCoeff13VI}

The Black Saturn line element~\cite{EF} reads
 \bea
  \nonumber
  ds^2 =
  -\frac{H_y}{H_x} \Big[dt + \Big(\frac{\omega_\psi}{H_y}+q\Big) \,
  d\psi \Big]^2
  + H_x \bigg\{ k^2 \, P \Big( d\rho^2 + dz^2 \Big)
       + \frac{G_y}{H_y} \, d\psi^2 + \frac{G_x}{H_x}\, d\varphi^2 \bigg\} \, ,
 \label{SaturnMetric}
\eea
where $k$, $q$ are real constants.
The contravariant components of the metric tensor are $g^{\psi\psi}=H_y/(H_x G_y)$,
$g^{\rho\rho}=g^{zz}=1/g_{\rho\rho}$, $g^{\varphi\varphi}=1/g_{\varphi\varphi}$ and
\bel{25III10.1x}
 g^{tt}=-\frac{H_x}{H_y}+\frac{H_y}{H_xG_y}\left(\frac{\omega_\psi}{H_y}+q\right)^2
 =  - \frac{g_{\psi\psi}}{G_y} \;,\quad
 g^{t\psi}=-\frac{H_y}{H_xG_y}\left(\frac{\omega_\psi}{H_y}+q\right)
 \;.
\ee
If we let
$$ \mu_i :=\sqrt{\rho^2
 + (z-a_i)^2}- (z-a_i)
  \;,
$$
where the $a_i$'s are real constants, then
\bea
  G_x = \frac{\rho^2\mu_4}{\mu_3\, \mu_5} \, ,
\eea
\bea
  P =  (\mu_3\, \mu_4+ \rho^2)^2
      (\mu_1\, \mu_5+ \rho^2)
      (\mu_4\, \mu_5+ \rho^2) \, ,
      \label{eqn:defP}
\eea
\bea
   H_x &=& F^{-1} \,
   \bigg[ M_0 + c_1^2 \, M_1 + c_2^2\,  M_2
   +  c_1\, c_2\, M_3 + c_1^2 c_2^2\, M_4 \bigg] \, , \label{eqn:defHx}\\[2mm]
\nonumber    H_y &=& F^{-1} \,
   \frac{\mu_3}{\mu_4}\,
   \bigg[ M_0 \frac{\mu_1}{\mu_2}
   - c_1^2 \, M_1 \frac{\rho^2}{\mu_1\,\mu_2}
   - c_2^2\,  M_2 \frac{\mu_1\,\mu_2}{\rho^2}
   +  c_1\, c_2\, M_3
   + c_1^2 c_2^2\, M_4 \frac{\mu_2}{\mu_1} \bigg] \, ,\\&&
\eea
where $c_1$ and $c_2$ are real constants, and
\bea\nonumber
  M_0 &=& \mu_2\, \mu_5^2 (\mu_1-\mu_3)^2 (\mu_2-\mu_4)^2
   (\rho^2+\mu_1\,\mu_2)^2(\rho^2+\mu_1\,\mu_4)^2
   (\rho^2+\mu_2\,\mu_3)^2 \, ,\\&& \\[2mm]
\nonumber M_1 &=& \mu_1^2 \, \mu_2 \, \mu_3\, \mu_4 \, \mu_5 \,
\rho^2\,
  (\mu_1-\mu_2)^2 (\mu_2-\mu_4)^2(\mu_1-\mu_5)^2
  (\rho^2+\mu_2\,\mu_3)^2  \, ,\\&& \\[2mm]
\nonumber
  M_2 &=& \mu_2 \, \mu_3\, \mu_4 \, \mu_5 \, \rho^2\,
  (\mu_1-\mu_2)^2 (\mu_1-\mu_3)^2
  (\rho^2+\mu_1\,\mu_4)^2(\rho^2+\mu_2\, \mu_5)^2  \, ,\\&& \\[2mm]
  M_3 &=& 2 \mu_1 \mu_2 \, \mu_3\, \mu_4 \, \mu_5 \,
  (\mu_1-\mu_3) (\mu_1-\mu_5)(\mu_2-\mu_4)
  (\rho^2+\mu_1^2)(\rho^2+\mu_2^2) \nonumber \\[1mm]
  &&\hspace{3cm} \times
  (\rho^2+\mu_1\,\mu_4)(\rho^2+\mu_2\, \mu_3)
  (\rho^2+\mu_2\, \mu_5)  \, ,\\[2mm]
  M_4 &=& \mu_1^2 \, \mu_2\, \mu_3^2 \, \mu_4^2 \,
  (\mu_1-\mu_5)^2
  (\rho^2+\mu_1\,\mu_2)^2(\rho^2+\mu_2\, \mu_5)^2  \, ,
\eea and \bea
  F &=& \mu_1\, \mu_5\,  (\mu_1-\mu_3)^2(\mu_2-\mu_4)^2
  (\rho^2+\mu_1\,\mu_3)
  (\rho^2+\mu_2\,\mu_3)
  (\rho^2+\mu_1\,\mu_4) \nonumber\\
  && \hspace{1cm} \times
  (\rho^2+\mu_2\,\mu_4)
  (\rho^2+\mu_2\,\mu_5)
  (\rho^2+\mu_3\,\mu_5)
  \prod_{i=1}^5 (\rho^2+\mu_i^2) \, .
\eea
Furthermore,
\bea
  G_y = \frac{\mu_3\, \mu_5}{\mu_4} \, ,
\eea
and the off-diagonal part of the metric is governed by
\bea
  \omega_\psi
  &=&
  2 \frac{
     c_1\, R_1\, \sqrt{M_0 M_1}
    -c_2\, R_2\, \sqrt{M_0 M_2}
    +c_1^2\,c_2\, R_2\, \sqrt{M_1 M_4}
    -c_1\,c_2^2\, R_1\, \sqrt{M_2 M_4}
  }
  {F \sqrt{G_x}} \, .
 \nonumber
\\
 \label{eq:offdiag}
\eea
Here $R_i = \sqrt{\rho^2 + (z-a_i)^2}$. We note that the square
roots in \eqref{eq:offdiag} are an artifact, in the sense that
the functions
$$
 {\frac{M_0M_1}{G_x}}\;,
 \quad
  {\frac{M_0M_2}{G_x}}\;,
 \quad
 {\frac{M_1M_4}{G_x}}\;,
 \quad \mbox{and} \
  {\frac{M_2M_4}{G_x}}
$$
can be checked to be complete squares, which implies that their
square roots can be rewritten as rational functions of the
$\mu_i$'s, $\rho^2$, and of the free constants appearing in the
metric.

The determinant of the metric reads
\bel{2X.2}
 \det g_{\mu_\nu} = -\rho^2 H_x^2 k^4 P^2
 \;.
\ee

\subsection{The parameters}

Here we summarise the restrictions imposed in~\cite{EF} on
various parameters appearing in the metric. The parameters
$a_i$ are ordered as
\beal{21IV.41.1}
 a_1\le a_5 \le a_4 \le a_3 \le a_2
 \;,
\eea
but throughout this paper we assume that the inequalities are
\emph{strict}.

Boundedness of $g_{tt}$ near $a_1$ leads either to $c_2=0$ or
to
\bel{21VII.4x}
 c_1=\pm\sqrt{\frac{2( a_3-a_1  ) ( a_4-a_1  )}{a_5  - a_1 }}
 \;.
\ee
This last condition follows also from the requirement of
boundedness of $g_{\psi\psi}$  near $a_1$ when $c_2=0$, and
thus \eq{21VII.4x} needs to be imposed in all cases. A choice
of orientation of $\psi$ leads to the plus sign.

From Table \ref{T18VII.0}, continuity of the metric at
$\{\rho=0,\ z<a_1\}$ leads to the condition
\bel{18VII.1} k = \frac{2 (a_{1}-a_{3})  (a_{2}-a_{4})   }
    {  2 (a_{1}-a_{3}) (a_{2}-a_{4})+(a_{1}-a_{5}) c_{1} c_{2}  }
    \;,
\ee
which can be checked to be  finite when the  value  of $c_1
c_2$ is inserted.

Asymptotic flatness requires
$$
 q=\frac{2 c_2\kappa_1}{
 2\kappa_1 - 2\kappa_1\kappa_2 + c_1 c_2\kappa_3}
 \;,
$$
as well as
$$k=-\frac{2\kappa_1 (-1 +\kappa_2)}{\sqrt{(-2\kappa_1 (-1 +\kappa_2) +
    c_1 c_2\kappa_3)^2}}
    \;,
$$
where
$$
 \kappa_i:= \frac{a_{i+2}-a_1}{a_2-a_1}
 \;,
$$
which can be checked to be consistent with \eqref{18VII.1}.

A conical singularity on the rotation axes of
$\partial_\varphi$ is avoided if
$$
c_2= \sqrt 2 ( a_4-a_2  ) \frac{\pm(a_1  - a_2  ) (a_3   - a_4  ) + \sqrt{(a_1  - a_3  ) ( a_4-a_2  ) (a_2   - a_5 ) (a_3   - a_5 )}}
{\sqrt{(a_1  - a_4  ) (a_2   - a_4  ) (a_1  - a_5 ) (a_2   - a_5 ) (a_3   - a_5 )}} \;.
$$

\section{Numerical evidence for stable causality}
 \label{nr}

In this Appendix we present numerical results that support the
conjecture that $g_{\psi \psi }$ is positive away from points
where $\partial_\psi$ vanishes. Regions where  $g_{\psi \psi }$
vanishes or becomes negative contain closed causal curves. On
the other hand, the conjecture implies stable causality of the
domain of outer communications, see Section~\ref{ss16IV10.1}.

While our numerical analysis indicates very strongly that
$g_{\psi \psi }$ is never negative in the region of parameters
of interest, it should be recognized that the evidence that we
provide concerning null orbits of $\partial_\psi$ is less
compelling.

The metric component $g_{\psi \psi }$ is a complicated function
of $\rho$, $z$ and the five parameters $a_{i=1,\dots,0}$. This
function is sufficiently complicated in the general case that
there appears to be little hope to prove non-negativity
analytically. We gave a complete analytic solution of the
problem in Section~\ref{Schoicec2zero} only for $c_2=0$. In
general, we turn to numerical analysis. The idea is to find an
absolute minimum of  $g_{\psi \psi }$.

The original phase-space of this minimization problem is seven
dimensional. One may use translation symmetry of Black Saturn
solution to reduce the dimension by one. We do this via the
choice $a_1=0$. Next choosing $a_5-a_1$ as a length unit leads
us to a five dimensional minimization problem. Our five
variables are $\rho$, $z$, $d_{45}$, $d_{34}$, $d_{23}$, where
$d_{ij}=a_i-a_j$. All of them are real and in addition
$\rho\geq 0$, $d_{ij}>0$.

The minimization procedure starts at a random initial point and
goes towards smaller values of $g_{\psi \psi }$. For general
$\rho\geq 0$ we use an algorithm  with gradient --- the so
called Fletcher-Reeves conjugate gradient algorithm. The limit
$\rho\rightarrow 0$ is non-trivial, therefore it has to be
studied separately. In this case, the values of the metric
functions are given by different formulas for different ranges
of $z$ coordinate. The expressions for the gradients are huge
and we did not succeed in compiling a C++ code with these
definitions. Therefore, for $\rho=0$ we use the Simplex
algorithm of Nelder and Mead. This algorithm does not require
gradients. Both algorithms are provided by the GNU Scientific
Library \cite{GSL}.

The minimisation procedure stops when the computer has attained
a local minimum by comparing with values at nearby points, or
when the minimizing sequence of points reaches the boundary of
the minimization region (coalescing $a_i$'s).
All local minima found by the computer were located very near
the axis $\rho=0$, where the results were unreliable because of
the numerical errors arising from the divisions of two very
small numbers, and it is tempting to conjecture that
$g_{\psi\psi}$ has non-vanishing gradient with respect to
$(\rho, z, a_i)$ away from the axis, but we have not able to
prove that.

The numerical artefacts, just described, were filtered out as
follows: Each value of $g_{\psi\psi}$ at a local minimum, as
claimed by the  C++  minimisation procedure  was recalculated
in {\sc Mathematica}. If the relative error was bigger than
$10^{-6}$, then the point was classified as unreliable and
excluded from the data. In particular all points at which C++
claimed a negative value of $g_{\psi\psi}$ were found to be
unreliable according to this criterion.

\begin{figure}[th]
\begin{center}
\includegraphics[width=7cm,angle=-90]{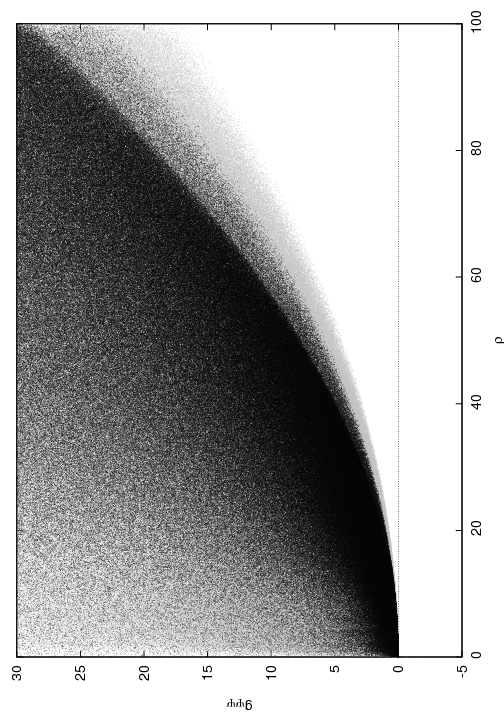}
\end{center}
\caption{The values of
$g_{\psi\psi}$ as a function of $\rho$ at the end of the minimization procedure;
this occurs
either at local minima, or at points where the minimizing sequence leads to coalescing
$a_i$'s. The three samples a), b), c) are presented
with different grey intensity (from low to high, respectively). The initial parameters
$(z, d_{ij})$ for the minimization procedure were randomly chosen,
uniformly distributed in the intervals
a) $z \in (-150,301)$, $d_{ij}\in (0,50)$,
b) $z \in(-150,226)$, $d_{ij}\in (0,25)$,
c) $z \in (-150,166)$, $d_{ij}\in (0,5)$.
For each sample, the minimum of $g_{\psi\psi}$ is proportional to $\rho^2$.}
\label{fig:plot1}
\end{figure}

Figure~\ref{fig:plot1} illustrates  a roughly quadratic lower
bound on
$$
 g_{\psi\psi}|_{ \rho\geq 0,
 z\in[-z_{\max{}},z_{\max{}}]}
  \;,
$$
with a slope depending  on the
collection $(z_{\max{}},d_{ij})$.

\begin{figure}[ht]
\begin{center}
\includegraphics[width=7cm,angle=-90]{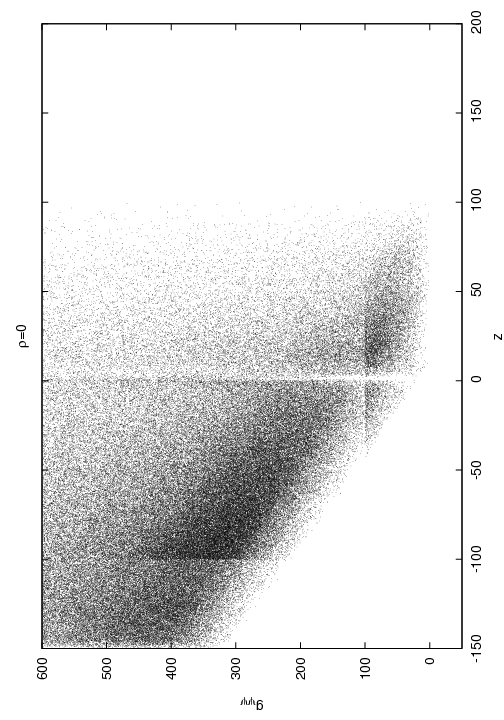}
\end{center}
\caption{The values of $g_{\psi\psi}$ for $\rho=0$ at the end of the minimization procedure;
this occurs at points where the minimizing sequence leads to coalescing $a_i$'s.
The initial parameters $(z, d_{ij})$ for the minimization procedure were randomly chosen,
uniformly distributed in the intervals $z\in(-150,301)$, $d_{ij}\in (0,50)$.}
\label{fig:plot1S}
\end{figure}

In Figure~\ref{fig:plot1S} one  observes a linear lower bound
on $g_{\psi\psi}|_{\rho=0}$ for $z<a_1$, with a slope
approximatively equal to $-2$ with our choice of scale
$a_5-a_1=1$.

The numerical results presented in this section support the
hypothesis that $g_{\psi\psi}$ is never negative in the region
of parameters of interest, vanishing only on the axis of
rotation $\{\rho=0 \;, \ z\ge a_2\}$.

\bigskip

\noindent{\sc Acknowledgements} PTC acknowledges useful
discussions with Christopher Hopper. We thank Marcus Ansorg,
Henriette Elvang, J\"org Hennig and Pau Figueras for useful
comments.

The research was partly carried out with the supercomputer
``Deszno" purchased thanks to the financial support of the
European Regional Development Fund in the framework of the
Polish Innovation Economy Operational Program (contract no.\ POIG.02.01.00-12-023/08).

The main part of our calculations was carried out with {\sc
Mathematica} together with the {\sc xAct} \cite{xAct} package.
We are grateful to Jos\'e Mar\'ia Mart\'in--Garc\'ia and
Alfonso Garc\'ia--Parrado for sharing their {\sc Mathematica}
and {\sc xAct} expertise.

\bibliographystyle{amsplain}
\bibliography{../references/hip_bib,%
../references/reffile,%
../references/newbiblio,%
../references/newbiblio2,%
../references/bibl,%
../references/howard,%
../references/bartnik,%
../references/myGR,%
../references/newbib,%
../references/Energy,%
../references/netbiblio}
%
\end{document}